\documentclass[twocolumn,showpacs,superscriptaddress, preprintnumbers , amsmath,amssymb,letterpaper]{revtex4-1}
\usepackage{lineno}
\usepackage{dcolumn}% Align table columns on decimal point
\usepackage{bm}% bold math
\usepackage{amsmath}
\usepackage{multirow}
\usepackage{graphicx}
\usepackage{float}
\usepackage[colorlinks=true]{hyperref}
\newcommand{\mean}[1]{\mbox{$\langle{#1}\rangle$}}

\begin{document}
\title{Spatial Control of Photoemitted Electron Beams \\
using a Micro-Lens-Array Transverse-Shaping Technique}
\author{A. Halavanau$^{1,2}$, G. Qiang$^{3,4}$, G. Ha$^{5}$,  E. Wisniewski$^{3}$,  P. Piot$^{1,2}$, J. G. Power$^{3}$, W. Gai$^3$, \\
$^1$ Department of Physics and Northern Illinois Center for Accelerator \& \\
Detector Development, Northern Illinois University, DeKalb, IL 60115, USA \\
$^2$ Fermi National Accelerator Laboratory, Batavia, IL 60510, USA \\
$^3$ Argonne Wakefield Accelerator, Argonne National Laboratory, Lemont, IL, 60439, USA\\
$^4$ Accelerator laboratory, Department of Engineering Physics, Tsinghua University, Beijing, China\\
$^5$ POSTECH, Pohang, Kyoungbuk, 37673, Korea}

\begin{abstract} 
A common issue encountered in photoemission electron sources used in electron accelerators is the transverse 
inhomogeneity of the laser distribution resulting from the laser-amplification process and often use of frequency up conversion in nonlinear crystals.
A inhomogeneous laser distribution on the photocathode produces charged beams with lower beam quality. In this paper, we explore the possible use of microlens arrays (fly-eye light condensers) to dramatically improve 
the transverse uniformity of the drive laser pulse on UV photocathodes. We also demonstrate the use of such microlens arrays to generate transversely-modulated electron beams and present a possible application to diagnose the properties of a magnetized beam. 

\end{abstract}
\date{\today}

\maketitle
\section{Introduction}
Photoemission electron sources are widespread and serve as backbones if an increasing number of applications 
including, e.g., high-energy particle accelerators, accelerator-based light sources, 
or ultra-fast electron diffraction setups. For a given photoemission electron
-source design, the electron-beam properties, and notably its brightness, are ultimately 
limited by the emission process and especially the initial conditions set by the laser
pulse impinging the photocathode. A challenge common to most applications is the ability 
to produce an electron beam with uniform transverse density. Non uniformities in the transverse electron-beam density can
lead to transverse emittance dilution or intricate correlations. 
Producing and transporting a laser pulse while preserving a homogeneous transverse density is challenging and 
has been an active area of work~\cite{siqili}. 
For instance, the ultraviolet (UV) laser pulses typically employed for photoemission 
from metallic or semiconductor cathodes requires the use of nonlinear conversion process 
to form the UV pulse from an amplified infrared (IR) pulse.  This frequency up-conversion 
mechanism often introduces transverse inhomogeneities owing to the nonlinearity of the conversion process. 

In this paper, we investigate an alternative simple technique capable of controlling the 
transverse shape of a UV laser pulse. The technique employs microlens arrays (MLAs) 
to directly homogenize the UV laser pulse. MLAs are commonly employed as optical
homogenizers for various applications~\cite{mlaabcd,dickey2000laser,deOliveira}. In 
addition to its homogenizing capability, we also demonstrate that the MLA-based technique can also 
produced a periodic transverse pattern  that can form a two-dimensional array of transversely-segmented beams. 
Such type of beams could find application in beam-based diagnostics of accelerator, single-shot 
quantum-efficiency map measurement, and coherent light sources in the THz regime 
or at shorter wavelength~\cite{Shibata:1994,graves}. 

In this paper, after briefly summarizing the principles of the MLA setup, we
demonstrate its possible use to homogenize the ultraviolet (UV) 
laser spot of the photocathode drive laser. We especially establish the usefulness 
of MLAs to control the electron beam distribution in a series of experiments carried 
out at the Argonne Wakefield Accelerator (AWA) facility~\cite{manoel}.

%. We performed electron beam emittance measurement and demonstrated the possibility of generation mutli-beam particle distribution using MLAs. 
%Finally, we demonstrate the usefulness of this technique by presenting a method of measuring canonical angular momentum (CAM) of the electron beam,
%introduced via non-zero magnetic field at the photocathode. 

\section{Optical performances of the MLA}

Qualitatively, the principle of the MLA lies in redistributing the incoming 
light intensity across the light beam spot. Typically, MLAs are arranged in pairs. 
After passing through the MLA assembly, the light rays are collected by a ``Fourier" 
lens which focuses parallel rays from different light beamlets to a single point at the
image plane. Under proper conditions (distance to the Fourier lens and its focal length), 
the process leads to transverse homogenizing of the beam; see Fig.~\ref{drawing}. Therefore
the MLA homogenization scheme is rather simple and appealing in the context of photocathode drive lasers.  

Alternatively, imaging the object plane of the single microlenses in the MLA with
a ``Fourier" lens produces  a set  of optical beamlets arranged as arrays (with a pattern
mimicking the microlens spatial distributions).

\subsection{ABCD formalism}

\begin{figure}[b]
\begin{center}
 \includegraphics[width=1\linewidth]{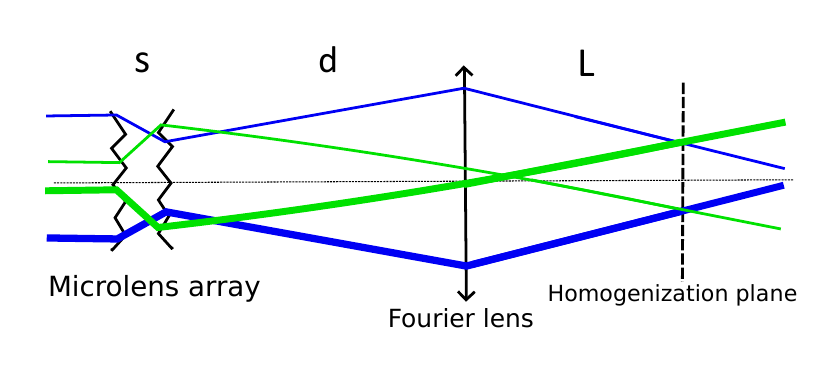}
 \caption{\label{drawing} Schematics of the microlens array configuration. 
 Initial intensity fluctuations in the beam (thin/thick ray) becomes
evenly distributed at the homogenization plane.}
\end{center}
\end{figure}

We first analyze the typical MLA setup diagrammed in Fig.~\ref{drawing} to derive a few salient 
features relevant to homogenization using the {\sc abcd} formalism~\cite{mlaabcd}. We consider an
initial ray to be characterized by the vector
$(x_0, x'_0)$, where $x_0$ and $x'_0\equiv\frac{dx_0}{dz}$ are 
respectively the initial ray position and divergence (here, $z$ represents the 
path-length along the optical transport). As a simple example, we consider a rectangular 
array of microlens in the $(x,y)$ plane with an equal pitch in both transverse directions. Using the  {\sc abcd} formalism, and considering that the ray is within the aperture $\rho$ of  the lens with center located at $(x=mp,y=np)$, we can describe the  MLA with the linear transformation 
\begin{eqnarray}
\left( \begin{array}{c}
x_{1}-mp  \\
x'_{1}  \end{array} \right) =  \left( \begin{array}{cc}
1 & 0  \\
-1/f_2 & 1  \end{array} \right) \left( \begin{array}{cc}
1 & s  \\
0 & 1  \end{array} \right) \times \\ \times \left( \begin{array}{cc}
1 & 0  \\
-1/f_1 & 1  \end{array} \right) \left( \begin{array}{c}
x_0-mp  \\
x'_0  \end{array} \right) \nonumber, 
\end{eqnarray}
where $(x_{1},x'_{1})$ is the ray vector after two MLA plates, $s$ is the spacing between two plates, $p$ is the array pitch, $f_1$
and $f_2$ are the focal lengths of the first and second microlens respectively. It 
should be pointed out that the ray initial and final 
coordinate satisfy $\sqrt{(x_0-mp)^2+(y_0-np)^2}\le \rho$ where $n$ and $m$ are integers 
that specify the position of each micro-lens in term of the pitch. Then, the 
output ray from the MLA setup can be further propagated up to the homogenization plane as 
\begin{eqnarray}
\left( \begin{array}{c}
x_{h}  \\
x'_{h}  \end{array} \right) =  \left( \begin{array}{cc}
1 & L  \\
0 & 1  \end{array} \right) \left( \begin{array}{cc}
1 & 0  \\
-1/F & 1  \end{array} \right) \left( \begin{array}{cc}
1 & d  \\
0 & 1  \end{array} \right) \left( \begin{array}{c}
x_1  \\
x'_1  \end{array} \right), 
\end{eqnarray}
where $(x_{h},x'_{h})$ is the ray vector at the homogenization plane, $d$ the 
distance between the Fourier lens and the MLA, $F$ the focal length of the Fourier 
lens and $L$ is the distance to the homogenization plane. 

From the formalism above one can deduce a few useful expressions. First, 
we consider the case when the two MLAs are identical ($f_1=f_2=f$) and located in the 
object plane of the Fourier lens ($L=F$). We further assume that there is no cross-talk 
between the microlens and their transformation only affects rays within a finite aperture 
smaller than the array pitch $\sqrt{(x_0-mp)^2+(y_0-np)^2}\le p/2$. Under these assumptions, 
we find the diameter of the image at the homogenization plane to be  
\begin{eqnarray}
\label{hom_spot}
 D_h \approx \frac{F p}{f^2}(2f-s)  
\end{eqnarray}
in the limit of small ray divergence (as indicated by the independence of the equation on $d$).
For practical purposes, we also calculate the diameter of the beam at the Fourier lens plane to be 
\begin{eqnarray}
\label{lens_apt}
 A_F \approx \frac{d p}{f^2}(2f-s).
\end{eqnarray}
The latter equation is useful to estimate the required aperture. 

In practice, the assumption $L=F$ might be challenging to satisfy. In such cases, the following expression is useful
to find the beam size at a given location $L$ with respect to ``Fourier'' lens:
\begin{eqnarray}
D (L) \approx \frac{p L}{f^2}(2f-s)+\frac{d p (2f-s)}{f^2}\frac{F-L}{F}. 
\end{eqnarray}

If  $L\approx F$ the resulting image remains homogenized due to the finite size of the Airy disk.
Moving away from the focal plane increases the density modulations and eventually yield an array of beamlets. 

\subsection{Optical transport design}

Photoinjector setups often incorporate relatively long (multi-meter scales) optical
transport lines. The optical lines include transport from the laser room to the photoinjector 
enclosure (generally performed in the air or in moderate vacuum pipe) 
and the injection in the ultra-high-vacuum accelerator beamline up to the
photocathode. Consequently, it is necessary to devise an optical transport line 
capable of imaging the homogenized laser profile on the photocathode surface. 
A commonly-used imaging setup, known as $4f$-imaging, is challenging to implement 
in the present case as it would require some of the lenses to be located in the vacuum chamber, 
as the ``imaging'' plane has to be much farther downstream than the ``object'' plane upstream.

However, imaging can be achieved in numerous ways while accommodating 
the various constraints related to MLAs (limited apertures, 
available focal lengths, etc...). To construct the appropriate optical line 
we impose the vector of a ray in the homogenization plane $(x_h,x'_h)$ to be 
transported to a downstream imaging plane $(x_I,x'_I)$ via 
\[\left( \begin{array}{c}
x_I  \\
x'_I  \end{array} \right) = \mathbf{M} \left( \begin{array}{c}
x_h   \\
x'_h \end{array} \right), \quad \mbox{with~~}\mathbf{M}=\left( \begin{array}{cc}
{\cal M} & 0  \\
0 & 1/{\cal M}\end{array} \right),\]
where the magnification ${\cal M}$ is set to 1 for one-to-one imaging. The latter linear system yields four equations; an additional
constraint comes from the total length of the imaging transport. Therefore, the problem has 5 unknowns in total with some flexibility within available lenses.
Hence, it is possible to construct four-lens solution with distances between lenses as free parameters to make 
the corresponding system of linear equations well-defined.

\begin{figure}
\begin{center}
 \includegraphics[width=1\linewidth]{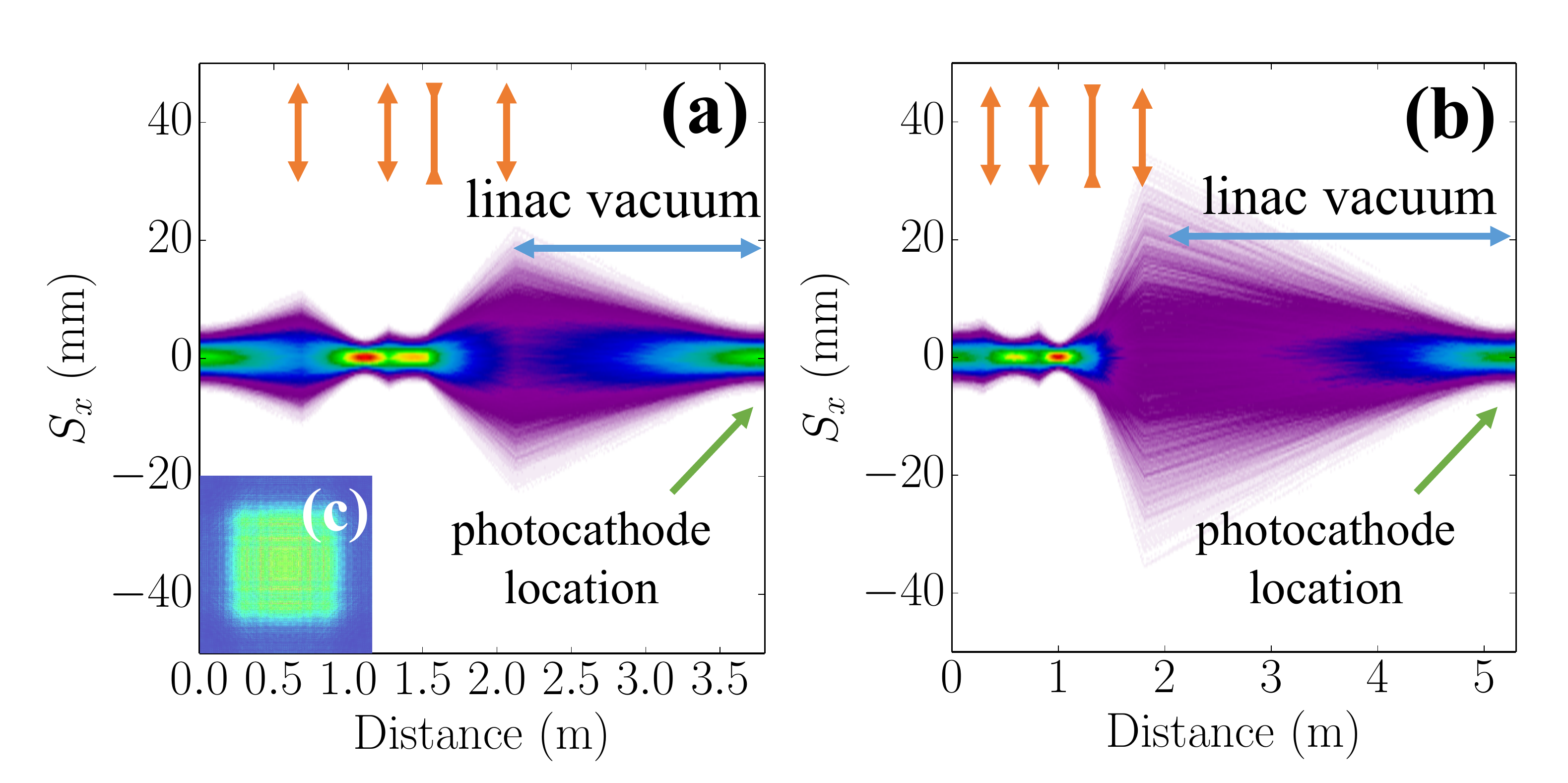}
 \caption{\label{image_transport} False color ray-tracing distribution of a four-lens
 optical line capable of imaging the homogenized beam on the photocathode surface. The configuration in (a)  and (b) correspond respectively to the AWA witness-beam and drive-beam electron-source setups.  The lenses type and locations are shown as red arrows. The inset (c) gives the intensity distribution simulated using the ectorial-diffraction program {\sc srw} for a $5 \times 5$ rectangular MLA.}
\end{center}
\end{figure}

The simulation of such a four-lens system was accomplished with a simple ray-tracing program where an 
initial set of optical ray were distributed according to a two-dimensional Gaussian 
distribution in the $(x,x')$ optical trace space. The optical layout of the laser 
transport downstream of the MLA is depicted in  Fig.~\ref{image_transport}(a): it 
includes four cylindrical-symmetric lenses, an optical window that allows for the laser 
beam to be injected in an ultra-high-vacuum area and an in-vacuum metallic mirror that 
direct the laser beam on the cathode surface. The resulting evolution of the beam size 
along the transport downstream of the MLA and up to the photocathode is display in Fig.~\ref{image_transport}(a,b) 
for the two RF-gun configurations available at the AWA facility. 
For both setups, the large beam size produced at the location of the last optical lens
demands a large-aperture lens. The beam size downstream of it gradually decreases until it reaches its target transverse 
size on the photocathode surface (8~mm rms). The in-vacuum mirror located close to the last optical transport lens
can be another limiting aperture of the optical system and generally results in beam losses. 
For the two cases reported in Fig.~\ref{image_transport} (a) and (b) 
the MLA-to-cathode transmission due to the finite geometric aperture, window transmission coefficient,
losses in the lenses and mirrors was computed to be 57\% and 43\%.

The designs presented in the Fig.~\ref{image_transport} were also simulated
 with the {\sc synchrotron radiation workshop} ({\sc srw} software~\cite{SRW} which is based 
 on Fourier optics and readily include a wave-propagation treatment of the laser transport; 
 see Fig.~\ref{image_transport} (c) inset.
 It confirmed that diffraction effects in the setup are negligible compared to
 transmission losses in the optical system. In the future, the established 
 numerical model of the MLA will be used for customizing the micro-lens profiles, arrangement, and pitch.
 It should be noted that linearized ABCD approach is sufficient to set up MLA system and full wave propagation
 simulations may be omitted in the beginning.
 
 Finally, transverse instabilities coming from shot-to-shot jitter in 
 the transverse distribution displayed in Fig.~\ref{noarray} (left), would result in 
 charge fluctuations if the laser beam is collimated by an iris upstream of the MLA. 
 To improve the stability of the laser intensity we introduced a two-lens beam reducer in front of the MLA. 

\subsection{Optical measurements\label{sec:opticalanalysis}}

To evaluate the performance of the proposed scheme, we use two MLA's on the 
photocathode drive laser of the AWA \cite{manoel}. The input UV ($\lambda = 248$~nm) laser pulse is obtained from frequency tripling of an amplified IR pulse originating from a Titanium-Sapphire (Ti:Sp) laser system. Downstream of the frequency tripler the UV pulse is further amplified in 
a two-pass excimer amplifier before transport to the accelerator vault. The setup shown in Fig.~\ref{drawing} was followed by the optical transport line shown in Fig.~\ref{image_transport}. A calibrated UV-sensitive screen with associated CCD camera mounted downstream of the setup directly provided a measurement of the achieved transverse distribution.  

To gain confidence in the performances of the MLA setup, we first investigated the impact of a non perfectly collimated incoming laser beam. 
As it can be inferred from Fig.~\ref{drawing}, the homogenization can still be achieved even 
if the incoming beam has a small divergence. There is a critical value of beam divergence $\tan \theta=p/2f$  that causes 
destructive interference after the MLA and results in light loss \cite{sussinfo}. 

The beam size provided by Eq. \ref{hom_spot} was used in the optical relay setup and Eq. \ref{lens_apt}
justified the aperture value of the ``Fourier'' lens. Overall, we have observed a 
good agreement with Eq. \ref{hom_spot} and Eq. \ref{lens_apt}. The calculated laser beam size was
within the aperture of all optical elements and latter was confirmed experimentally.

Note, that the Fourier lens in the experimental setup should be placed at the distance $D>F$ from the array, where $F$ is the focal length of the Fourier lens. 

\begin{figure}[hh]
 \includegraphics[width=1\linewidth]{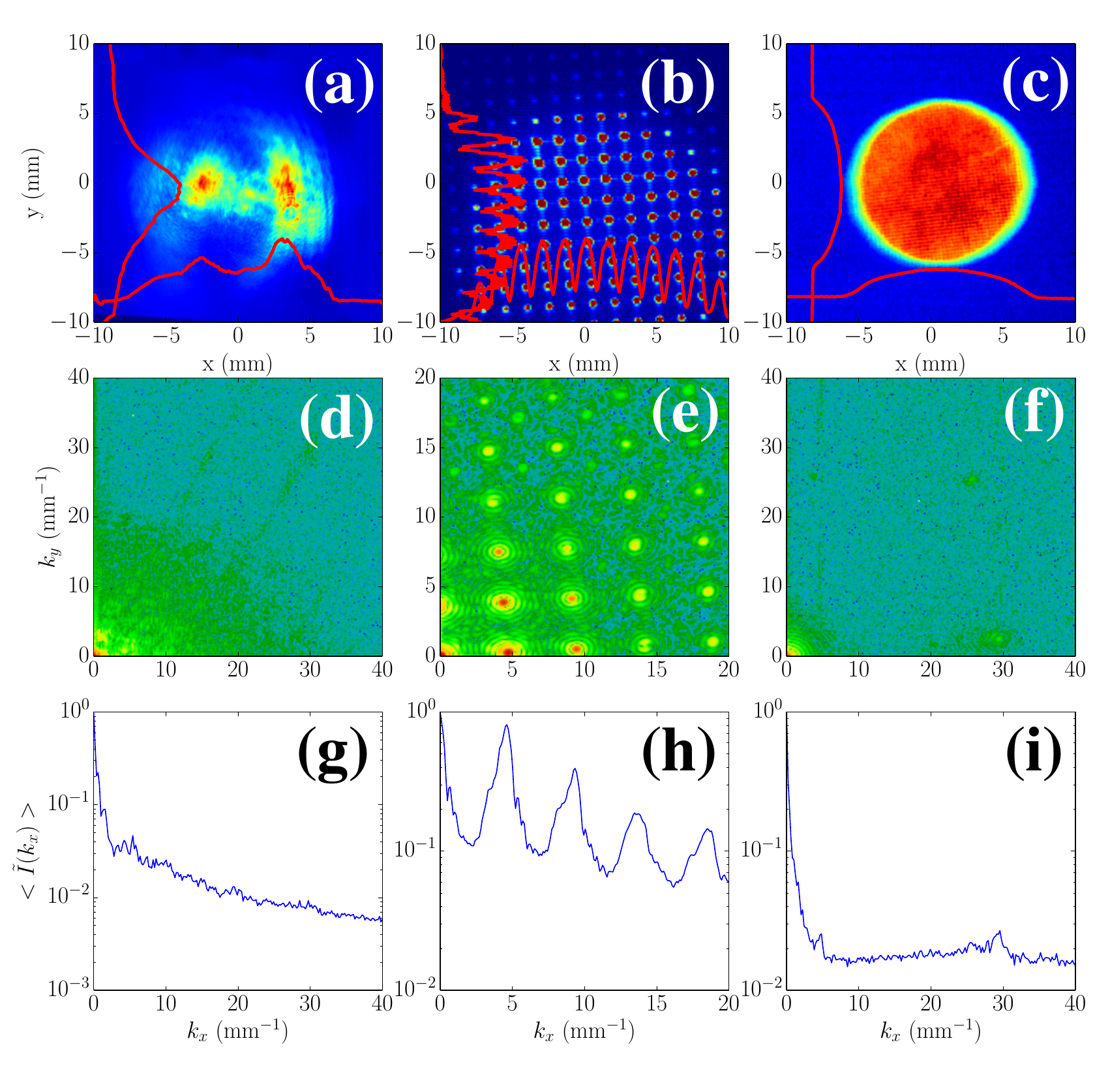}
 \caption{Measured UV laser without MLA (left column) and with MLA setup to
 produce beamlets (middle column) or as a homogenizer (right column). The upper, middle and lower rows respectively 
correspond to the laser transverse density distribution, its 2D FFT, 
and the projected  spectrum along the horizontal frequency $k_x$.\label{noarray}}
\end{figure}

The setup was also employed to demonstrate the homogenization process and quantify its performances. 
The nominal UV laser pulse was used as a starting condition; see Fig.\ref{noarray}(a).  The inhomogeneity of the transverse distribution can be quantified using
the spatial Fourier transform~\footnote{We note
another popular technique to quantify
the quality of an optical beam relies on 
the decomposition into Zernike's polynomials.
Our choice to use the two-dimensional Fourier transform was motivated by the need to 
use one figure of merit to quantify {\em both} the quality of 
the homogenized beam and to also parameterized the modulated beam.}.
Correspondingly, we consider the digitized image $I(x,y)$ associated to the transverse laser distribution and compute its two-dimensional (2D) Fourier transform $\tilde{I}(k_x,k_y)$ using the
fast-Fourier-tranform (FFT) algorithm available in the {\sc python}'s {\sc numPy} toolbox~\cite{python-numpy}. Here $k_x,k_y>0$ are the spatial wavenumbers respectively associated to the horizontal and vertical direction. In order to simplify the comparison we further introduce the one-dimensional Fourier transform $\tilde{I}_x(k_x)=\int_{0}^{+\infty} \tilde{I}(k_x, k_y) dk_y$ along the horizontal axis
[a similar definition hold for the vertical axis $\tilde{I}_y(k_y)$]. 
Figures~\ref{noarray} (d) and (g) respectively correspond
to the 2D Fourier transform and the projection along the horizontal wavenumber $k_x$ axis associated to the laser distribution displayed in Fig.~\ref{noarray} (a). It displays typical microstructures observed in previous runs at AWA, and the corresponding
spectrum displays some small modulations at low frequencies with most of the spectral content below $k_i<5$~mm$^{-1}$.
It should be noted that the excessive beam distortion observed in Fig.~\ref{noarray}(a) is the result of beam filamentation as the high-energy UV pulse propagates in the 20-m open-air optical transport system from the
laser room to the accelerator vault.

\begin{figure}[hh]
 \includegraphics[width=0.95\linewidth]{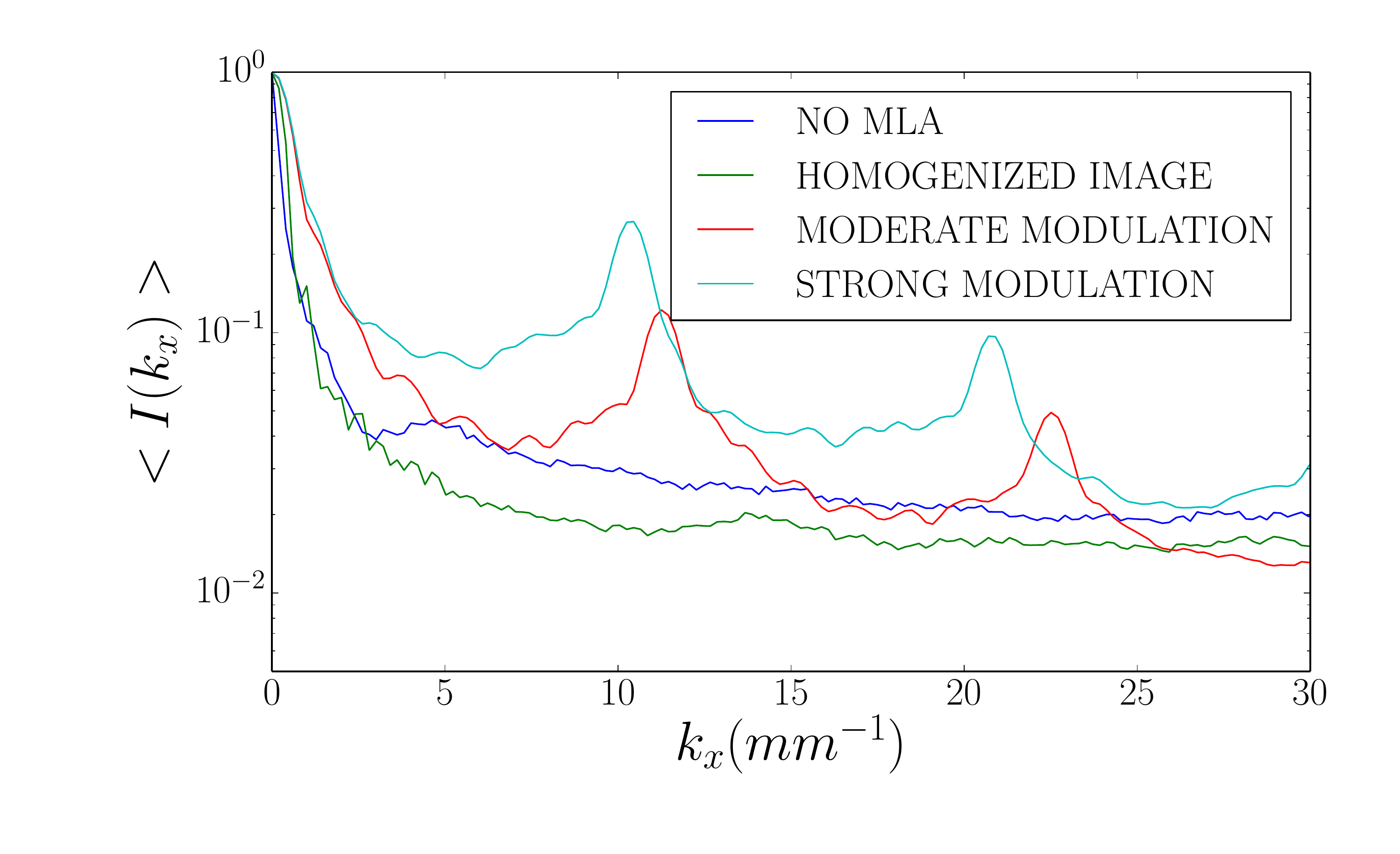}
 \caption{\label{MLA_FFT} FFT spectrum along the horizontal axis $\tilde I(k_x)$ for the different positions of the ``Fourier'' lens.
 The blue trace corresponds to no MLA case. 
 The green, red, and turquoise traces respectively correspond to the ``Fourier'' lens located at 
 250, 275 and 325 mm from the MLA array.} 
\end{figure}

When the MLA setup is configured to homogenize the beam [see Fig.~\ref{noarray}(c)], the Fourier transform indicates
that the although the low frequency modulations seen in the original beam are suppressed, high-frequency modulations 
are present for $k_x>12$~mm$^{-1}$. These modulations have a bunching factor on the order of $10^{-2}$ and 
correspond to very small modulation wavelength ($<$0.5 mm) barely observable on the distribution; see Fig.~\ref{MLA_FFT}.

Additionally, the MLA can be arranged to form a transversely-modulated laser distribution, the spectrum indicates 
a bunching factor at frequencies larger than the 
characteristic frequency associated to the total beam size; see Fig.~\ref{noarray}(b,e,h). We should point out 
that the non cylindrical-symmetric (square shaped) pattern transferred to the electron beam is eventually rotated 
due to the Larmor precession in the solenoidal lenses commonly surrounding RF guns. It is therefore important to 
mount the MLA assembling on a rotatable optical stage for remote control of the final pattern angle. 
Such an approach would decouple the downstream focusing (when solenoid are employed) and ensure the final
distribution does not have significant coupling between the two transverse degrees of freedom. Additionally, 
the fine control over the rotation of the final distribution could be used to select rotation angles with 
higher-order bunching to reach higher modulation frequencies, e.g., before injecting the beam in a 
transverse-to-longitudinal phase space exchanger to map the modulation into the temporal domain.

Figure~\ref{MLA_FFT} compares the projected horizontal Fourier spectra for four cases of MLA configurations. 
Each spectrum is obtained by averaging
five measurement taken after $f=250$ mm Fourier lens at  250 mm, 275 mm and 325 mm
to study the off-focal modulation and pattern formation. The latter Figure confirms that
in homogenization regime MLA setup significantly improves the image spectrum by 
suppressing the original low-frequency modulations in the beam. 

\begin{figure*}[tttttt]
\centering
\includegraphics[width=1\linewidth]{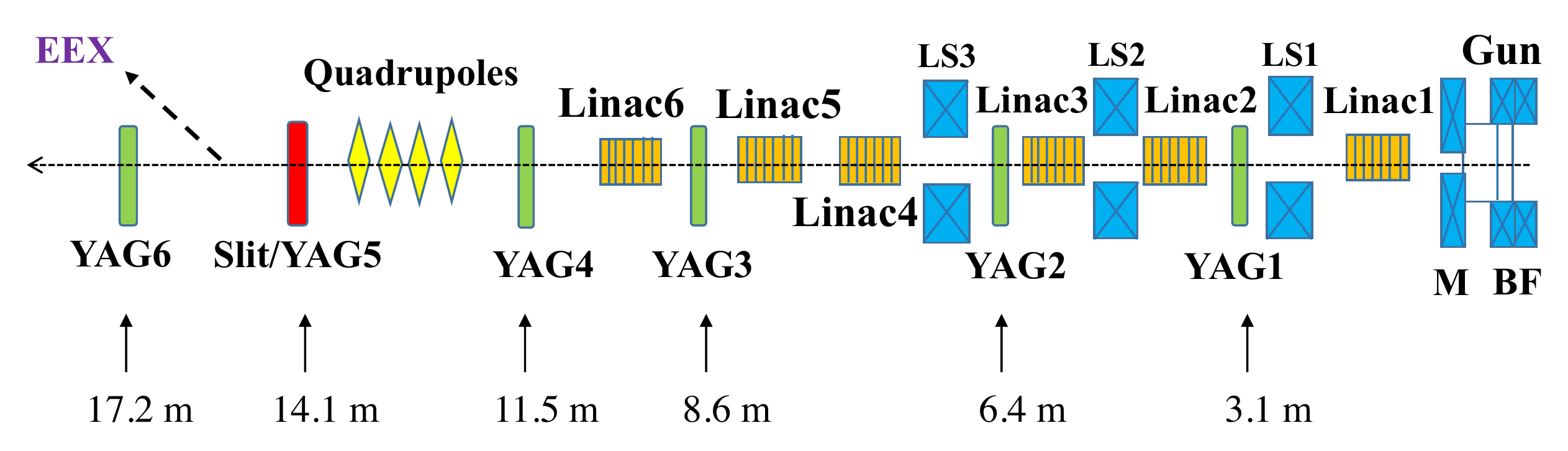}
\caption{\label{beamline}Overview of the AWA-DB  beamline only showing elements relevant to the performed experiment. 
Bucking-focusing 
(BF) and matching (M) solenoids were adjusted to image the beam on YAG screens. Linac solenoids (LS) and
quadrupoles were turned off during
the experiment. The positions of the YAG viewers are denoted in meters.
The energy gain of one accelerating cavity (linac) is 10~MeV. EEX label marks the separate double-dogleg
beamline for emittance exchange 
experiments}
\end{figure*}

Finally, we quantify the laser power loss in the devised setup. The MLA plates employed in our series of 
experiment do not have any UV anti-reflection (AR) coating, hence the power loss was $\sim 5$\% per
surface totaling $\sim 20$\% for the two MLAs. Additionally, the AR-coated UV lenses introduce a power 
loss of $\sim 2$\% per lens. In our optical setup the laser energy was measured to be
$4.2\pm 0.1$ and $2.5\pm 0.1$~mJ respectively
upstream and downstream of the MLA setup including the Fourier and four transport lenses.
Such a measurement indicates an energy transmission of $\sim 60$\% which could most likely be further
improved in an optimized setup. However given the UV laser energy available during our proof-of-principle 
experiment and the real-estate constraints we did not carry out such an optimization.

\section{Application of the MLA as a laser homogenizer}
\label{homogenizer}
The first set of experiments consisted in demonstrating the simple homogenization 
technique to improve the emittance of an accelerator.
The experiment was performed
in the AWA ``drive-beam" accelerator (AWA-DB) diagrammed in Fig.~\ref{beamline}. In brief, 
the transversely manipulated UV laser pulse impinges a high-quantum efficiency Cesium 
Telluride (Cs$_2$Te) cathode located in an L-band (1.3-GHz) RF gun to produce
7~MeV electron bunch. The electron bunches are then further accelerated in an
L-band normal conducting cavities up to 75 MeV. For a detailed description of 
the facility, the reader is referred to Ref.~\cite{manoel}. 
The RF gun is surrounded by three solenoidal lenses referred to as bucking,
focusing and main solenoids. The bucking and focusing solenoid have opposite polarity and are 
ganged to ensure the axial magnetic field on the photocathode vanishes. 
Several YAG:Ce scintillating screen (YAG in Fig.~\ref{beamline}) are available to
measure the beam transverse density along the accelerator beamline. 

\subsection{Beam dynamics simulations}

We carried out several simulations using the beam-dynamics program {\sc General Particle Tracker} (GPT)~\cite{GPT}
to  explore the impact of the MLA-homogenized beam on the resulting emittance.
Transverse inhomogeneities on the laser distribution at the photocathode surface are mirrored
on the photoemitted electron bunch distribution. These imperfection results in asymmetric space-charge 
forces and eventually yield phase-space dilution that ultimately degrade the beam emittances~\cite{FZhou}. 
Therefore the homogenized laser beam is expected to improve the beam transverse emittance.

\begin{figure}[hhh!!!]
\centering
 \includegraphics[width=0.97\linewidth]{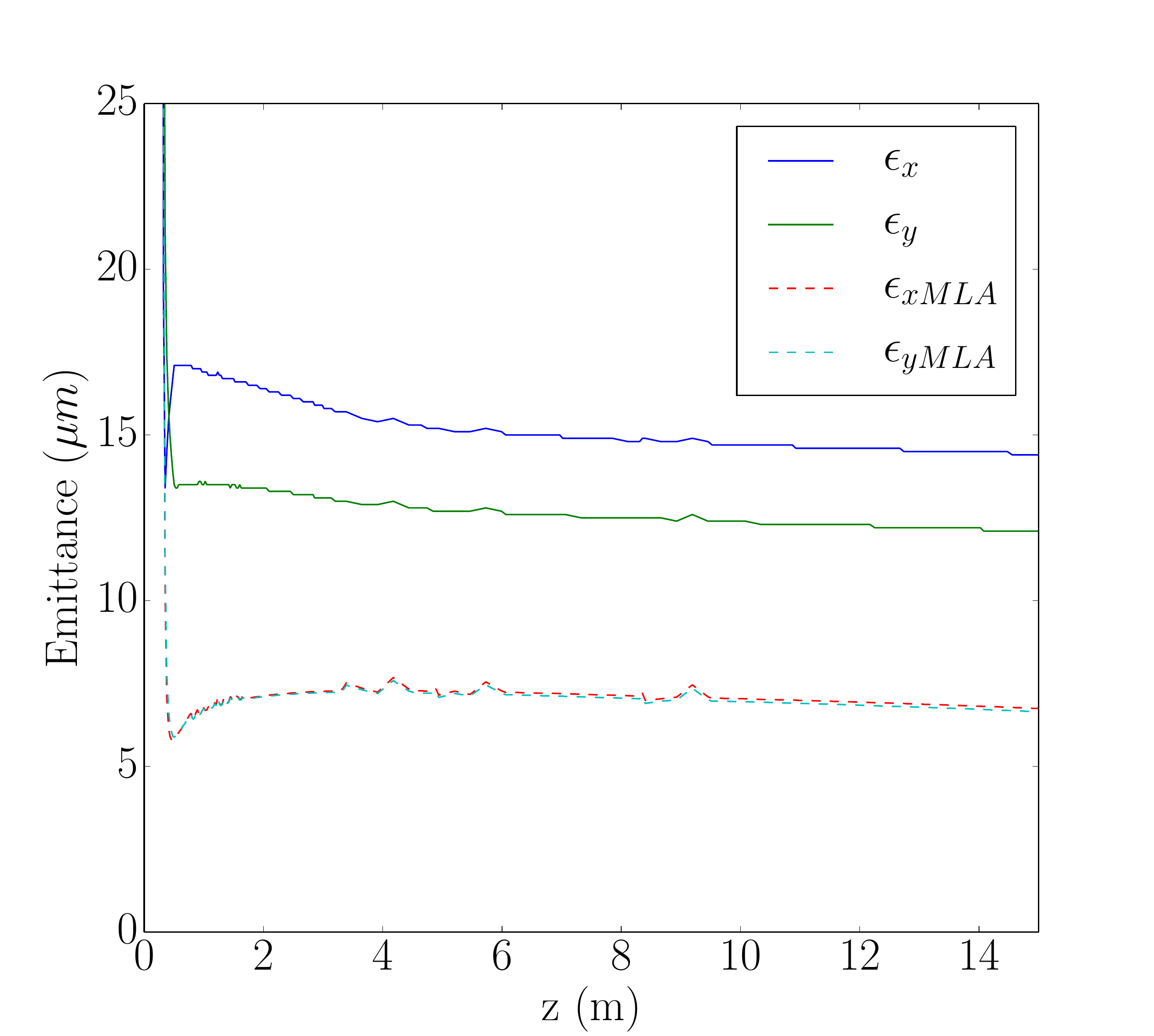}
 \caption{\label{beamemit2} Evolution of the transverse normalized emittances
 along the AWA-DB beamline simulated with {\sc{GPT}} for a 1-nC bunch. The simulations were performed 
 using as initial condition both the measured nominal (solid trace) and homogenized (MLA, dashed traces) 
 laser distributions.  The ordinate $z$ is the distance from the photocathode surface along the beamline.
 }
\end{figure}

The initial macroparticle distribution was produced using a Monte-Carlo generator 
using the measured transverse distribution of the laser similarly to Ref.~\cite{Rihaoui}.
The temporal laser distribution is taken to be Gaussian with RMS duration $\sigma_t=2.5$~ps, consistent
with streak camera measurements. The momentum of the macroparticle 
assumes an excess kinetic energy of 0.5~eV as typically 
considered for Cs$_2$Te cathodes~\cite{Flottmann:1997eu}. 
We considered the nominal and homogenized laser distribution respectively shown in Fig.~\ref{noarray}(a) and (c). 
To ensure a fair comparison, the total charge for both cases of distributions was set to 1~nC. Likewise, 
the RMS transverse sizes of the distribution was fixed to $\sigma_c=8$~mm along both the horizontal and vertical directions. 
The simulations demonstrate that the beam transverse emittances are reduced by a factor $\sim 2$ for the case of 
the homogenized laser distribution; see simulated row in Table \ref{emittable} and Fig.~\ref{beamemit2}.

\subsection{Transverse emittance measurements}

\begin{figure}[hhhhh!!!]
 \includegraphics[width=1.0\linewidth]{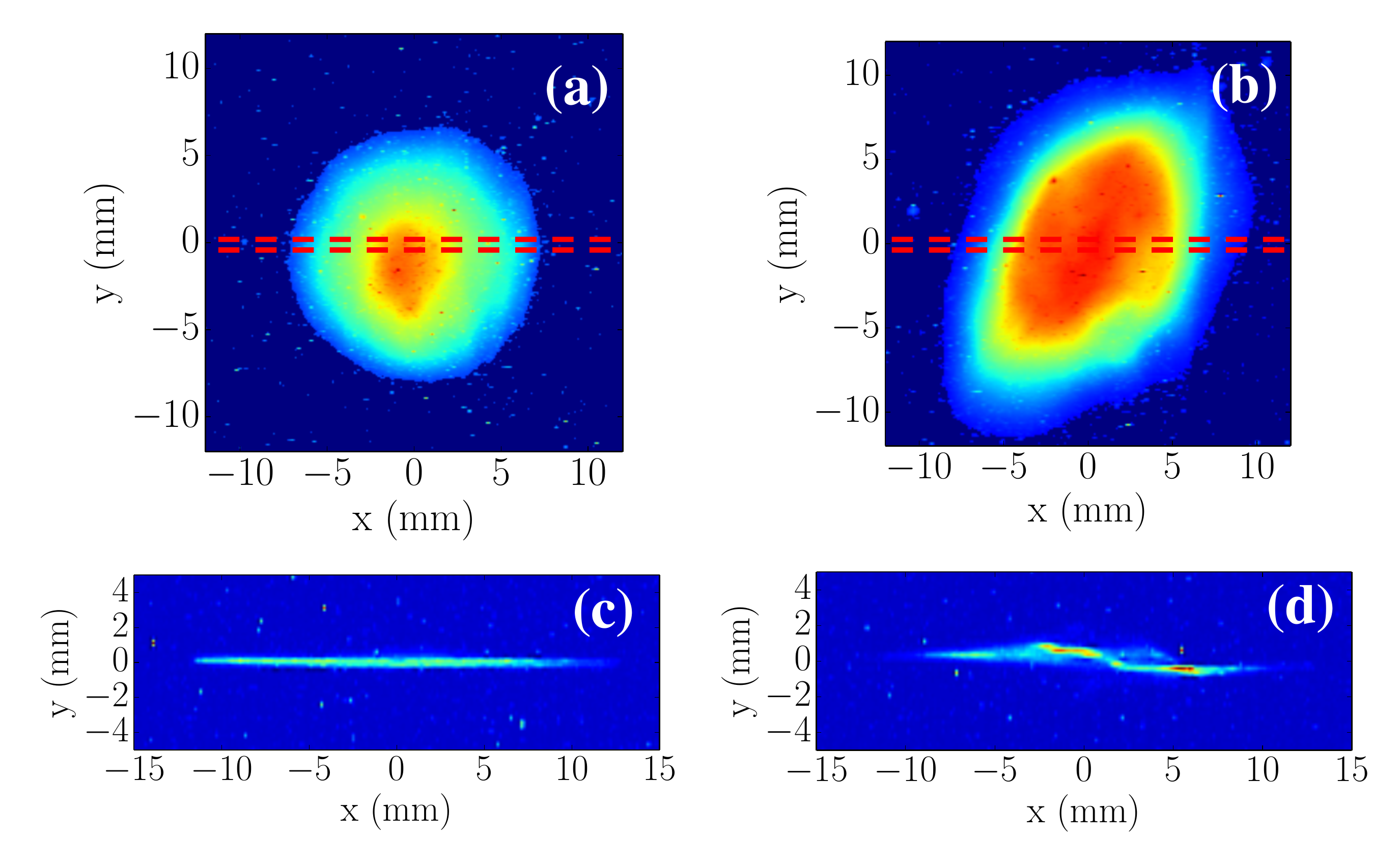}
 \caption{\label{mlabeam} Beam transverse distribution at YAG5
 (a,b) and associated distribution of the beamlet transmitted through a horizontal slit located at YAG5 
 location and measured at YAG6 (c,d). The set of images (a,b) [resp. (c,d)] corresponds 
 to the case when the MLA was inserted [resp. retracted] from the laser-beam path.
 The horizontal dash line in (a,b) represent the aperture of the slit.}
\end{figure}

The experimental verification of the benefits of homogenizing the laser
distribution was accomplished using the measured distribution of  Fig.~\ref{noarray}(a) and (c).
For the homogenized distribution displayed in Fig.~\ref{noarray}(c), a circular iris was used to clip
the laser distribution and ensure it had the same rms value as in Fig.~\ref{noarray}(a) 
$\sigma_c = 8\pm 0.2$~mm. The resulting electron beam was transported through 
the nominal AWA-DB beamline and accelerated to $p=48 \pm 0.5$~MeV/c. 
The corresponding electron-beam transverse distributions measured 
at YAG5 are compared in Fig.~\ref{mlabeam}(a,b). 
The distribution originating from the initial (inhomogenized) 
does not present any distortion except for beam asymmetric and having some $x-y$ coupling. 
In contrast, the homogenized distribution is cylindrically symmetric and does not show any coupling.
To further quantify the improvement we measured the beam vertical emittance using the slit technique.
A 100-$\mu$m wide horizontal slit was inserted at YAG5 and the transmitted beamlet
was observed 3.1 m downstream at YAG6 thereby providing the beam divergence $\sigma_y'$. 
Such a measurement together with the vertical beam-size measurement at YAG5
$\sigma_y$ provide an estimate of the normalized vertical emittance via $\varepsilon_y= \beta \gamma \sigma_y \sigma'_y$,
where $\beta \approx 1$ and $\gamma = 93.9$. It should be stressed that the reported emittance is the core emittance and does not
fully characterize the beam transverse phase space. Nevertheless this quantity provides a figure of merit to investigate the impact of the MLA on the beam quantity. 
The resulting beamlet distributions at YAG5/6 are shown in Fig.~\ref{mlabeam}(b,d)
and the measured divergence and resulting emittance compared in Tab.~\ref{emittable}.

The resulting emittances are comparable with the value simulated with GPT (Fig.~\ref{beamemit2})
and indicate a factor $\sim 2$ improvement for when the homogenized laser beam is employed. 
The relatively large error bars in Tab.~\ref{emittable} are due to hardware uncertainty (mostly the slit width). It should be noted that the errors between the
two measurements are correlated, i.e. the uncertainty leds to the upper (resp. lower) value for simultaneously
the ``MLA" and ``No MLA" measurements, thereby giving confidence, despite the large error bar on the emittance, that the emittance reduces
when the MLA is used to homogenize the laser beam.
It should finally be pointed out that the reported emittance were produced with a nominal setup of the  AWA-DB beamline, i.e. 
no emittance-minimization technique was attempted prior to the measurements.  
\begin{table}[]
\centering
\begin{tabular}{l c c c }\hline
\hline
%\multicolumn{1}{|c|}{\multirow{2}{*}
parameter  &    &  & units  \\
~& \multicolumn{2}{c}{experimental conditions} & ~ \\
  & No MLA  & MLA &    \\
\hline
~& \multicolumn{2}{c}{Simulation with GPT} & ~ \\
\hline
momentum $\mean{p}$                         &    48        &    48  &  MeV/c   \\ 
$\sigma_x$                         &    3.6       &    3.8     &  mm     \\ 
$\sigma_y$                         &    4.4        &    3.9     &  mm      \\ 
$\sigma'_x$                         &    4.6        &    1.7     & $\times 10^{-2}$ mrd      \\ 
$\sigma'_y$                         &    3.1        &    1.5     & $\times 10^{-2}$ mrd      \\ 
$\varepsilon_x$                         &    15.6         &    6.1  &  $\mu$m   \\ 
$\varepsilon_y$                         &    12.8         &    5.5    &  $\mu$m      \\ 
\hline
~& \multicolumn{2}{c}{Measurement} & ~ \\
\hline
momentum $\mean{p}$                         &    48 $\pm$ 0.5        &    48 $\pm$ 0.5 &  MeV/c   \\ 
$\sigma_x$                         &    4.4 $\pm$ 0.2        &    4.0 $\pm$ 0.2     &  mm     \\ 
$\sigma_y$                         &    5.2 $\pm$ 0.2        &    3.7 $\pm$ 0.2   &  mm      \\ 
$\sigma'_y$                         &     4.2 $\pm$ 1.3     &    3.3 $\pm$ 1.0    & $\times 10^{-2}$ mrd      \\ 
$\varepsilon_y$                                 &     20.5 $\pm$ 7.4               &     11.6 $\pm$ 4.3     &  $\mu$m        \\ 
\hline
\hline
\end{tabular}
\caption{\label{emittable}Comparison between measured 
and simulated beam parameters at YAG5 for $Q=1\pm 0.1$~nC. The experimental setup only allowed for the vertical normalized 
emittance to be measured. The parameters are all given as RMS quantities and corresponds to 
the distributions shown in Fig.~\ref{mlabeam}. }
%The quantity $\sigma_u$ (resp. $\sigma'_u$) corresponds to the RMS transverse beam size (resp. divergence) with $u=x,y$.}
\label{my-label}
\end{table}

\section{Production and transport of multi-beam arrays \label{multibeam}}
The application of the MLA setup investigated below is
the formation of transversely-segmented beams $-$ i.e.
consisting of an array of beamlets. Such a distribution could have a 
variety of applications such as described in Refs.~\cite{brainscan,japanesetherapypaper}.
Alternatively, the formed array could produce a transversely modulated
beam that could be injected in a transverse-to-longitudinal phase-space exchanger 
to yield a temporally-modulated beam~\cite{PhysRevSTAB.14.022801,PhysRevLett.105.234801}. 
The latter opportunity motivated the present work to demonstrate the generation and 
preservation of an array of beamlets up to the entrance of a
transverse-to-longitudinal phase-space exchanger installed in the AWA-DB~\cite{sunPAC07}
and recently employed for temporal shaping~\cite{PhysRevLett.118.104801}. 
Additionally, the multi-beam may serve as a beam-based diagnostic tool, e.g., to investigate nonlinearities 
of the externally-applied electromagnetic field or measure transfer matrices of beamline elements. 
In this section, we explore whether the beam transverse electron-beam modulation originating 
from the laser is preserved during the photoemission processes and low-energy acceleration in the RF gun. 

\subsection{Beam dynamics simulations}
Using the particle tracking codes {\sc gpt} and {\sc Impact-T} \cite{ImpactT}
we performed simulation of the AWA-DB RF-gun beam dynamics.
The preservation of the modulation is affected by space-charge forces
which play a dominant role in the beam dynamics in the vicinity of the cathode and in the RF gun.
Given the multi-scale nature of our problem, the 
space-charge forces are computed with a Barnes-Hut
(BH) algorithm \cite{Barnes:1986} available in {\sc gpt}. 
Similar algorithm was successfully tested in recent studies~\cite{Maxson2013,Halavanau:NIMA}. 

The measured transversely-modulated laser distribution at the photocathode location [similar to Fig.~\ref{noarray}(b)] was used to generate
the input macroparticle distribution for our numerical simulations. An initial 
intent was to probe whether the modulation could possibly be amplified via collective 
effects (e.g. implying transverse space charge modulations that will
eventually convert into energy modulations) or if they are simply
smeared out via thermal-emittance effect as the beam is photo-emitted.
 \begin{figure}[hhhhhhhh]
\centering
 \includegraphics[width=0.97\linewidth]{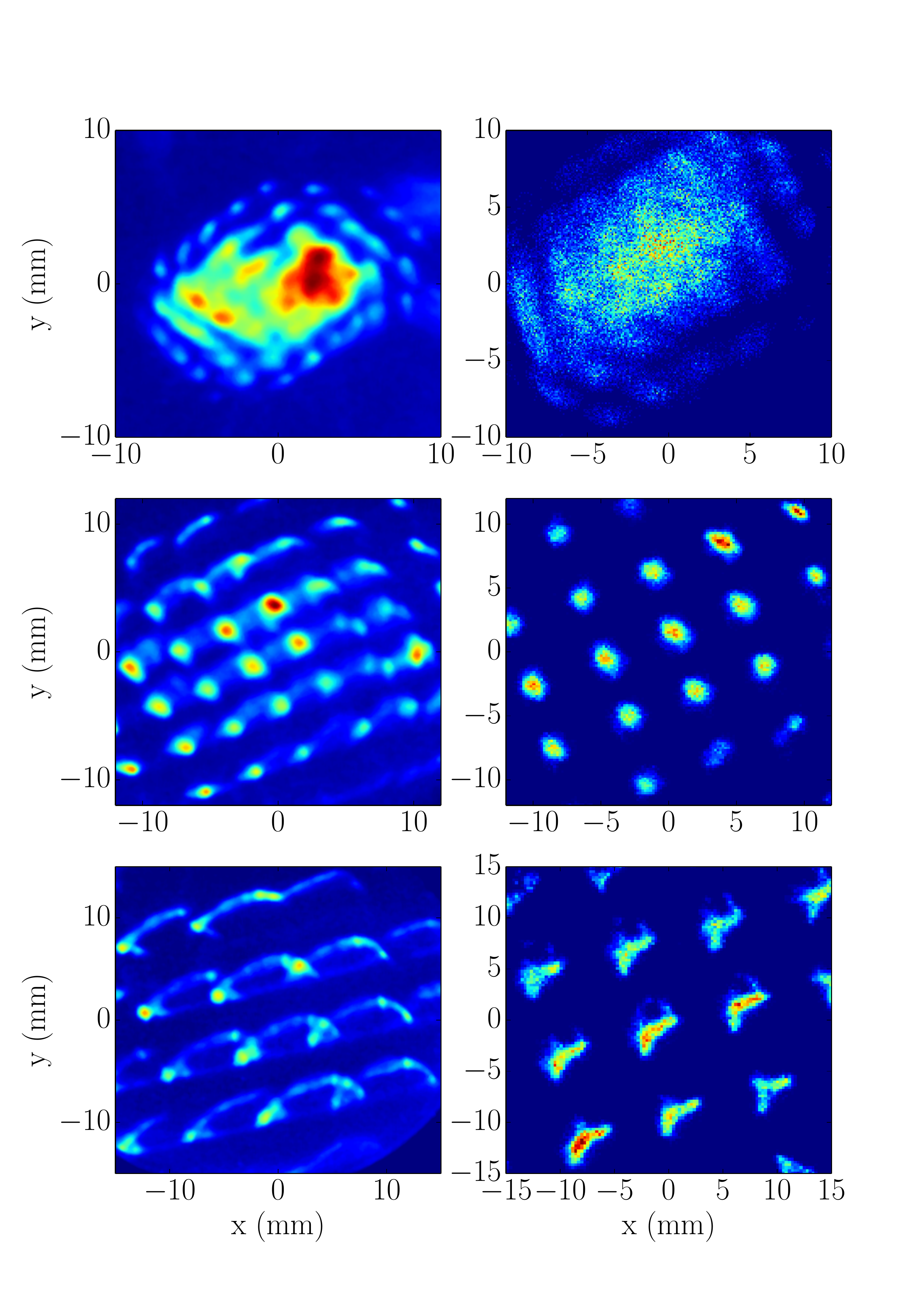}
 \caption{\label{simpat} Measured (left) and simulated (right) Q=100 pC electron-beam distribution at YAG1 when the UV laser pulse is modulated with the MLA array.
 The rows correspond to different matching-solenoid current settings of 215 A (upper row), 230~A (middle row) and 290~A (lower row). }
\end{figure}
The beamlet pattern was also used to ensure the {\sc gpt} model could reproduce
our experimental observation at low energy (RF-gun only). We compare the simulated
and measured transverse patterns for different matching-solenoid current settings
in Fig.~\ref{simpat}. The beam data and numerical simulations were recorded downstream
of the gun using YAG1 at a beam energy of $7 \pm 0.5$~MeV and bunch charge was 
set to 150~pC. The observed pattern rotation indicates that the Larmor angle which
depends on the beam energy and applied magnetic field agrees
qualitatively with the simulation; see Fig.~\ref{simpat}. The observed discrepancy is
caused by the uncertainties on the RF-gun field RF uncertainties in the RF-gun.

\subsection{Multi-beam formation downstream\\
 of the RF gun}
\begin{figure}
\centering
\includegraphics[width=1.0\linewidth]{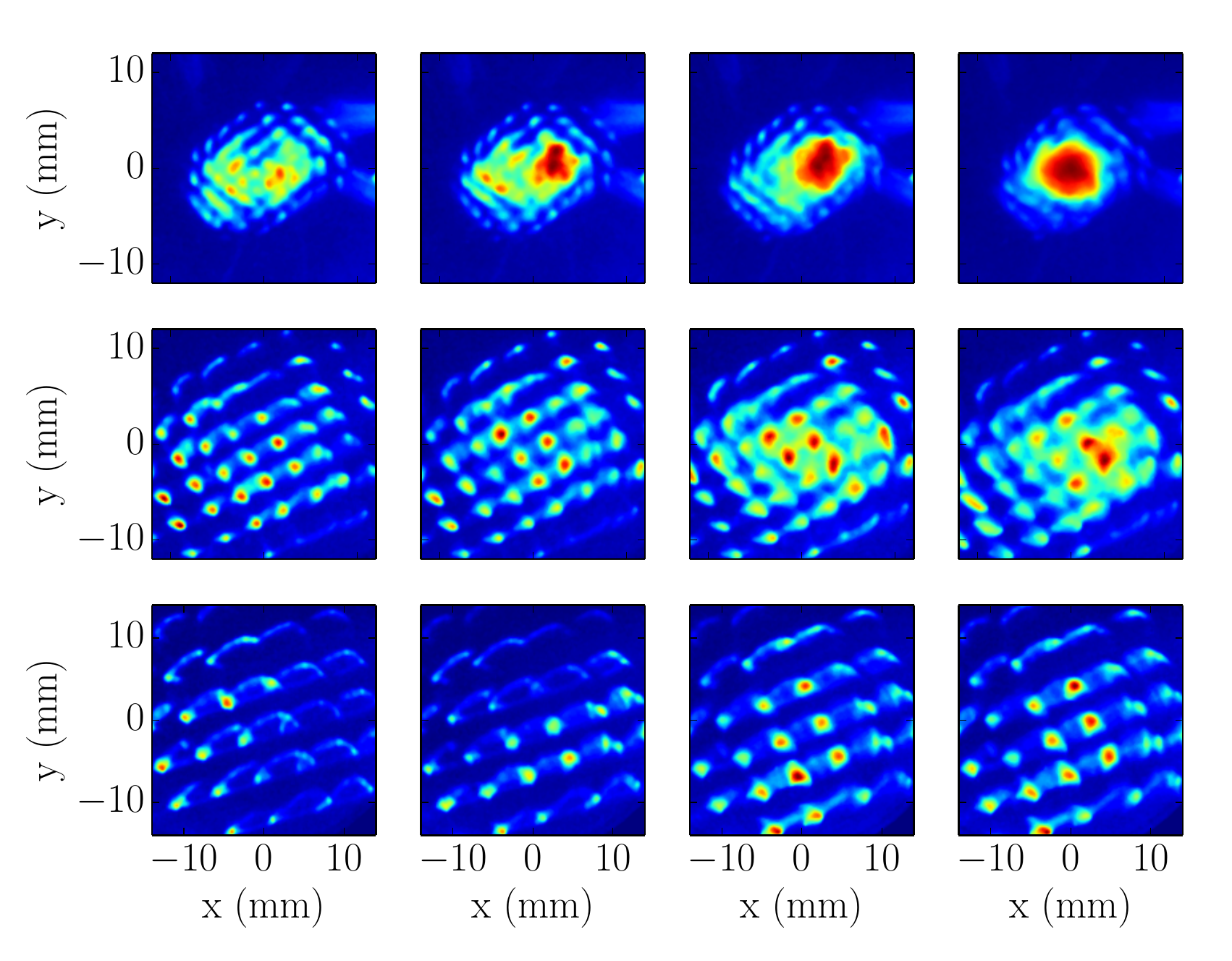}
\caption{\label{YAG1}False color measured 7 MeV electron beam patterns for various matching solenoid current setting and charge.
From left to right: Q=60pC, 80pC, 100pC, 120pC. The images 
from top to bottom correspond to matching-solenoid currents of 215, 240, and 270~A.}
\end{figure}

A subsequent experiment investigated the formation of a beamlet array downstream of the RF gun 
at an energy of $7 \pm 0.5$~MeV for various operating points of the photoinjector. The incoming laser 
spot size on the MLA array was chosen to yield an $8\times8$ beamlet array. The photoemitted electron
beam was observed on the YAG:Ce scintillating screen (YAG1 in Fig~\ref{beamline}) located at $z=3.1$~m from 
the photocathode surface. Figure~\ref{YAG1} displays a sequence of beam distribution recorded at
YAG1 for different settings of the focusing-bucking and matching solenoids. 
Note, that due to the surface space charge effects, 
the charge associated to each beamlet, and therefore the total maximum charge of the patterned beam, is limited. 
The total maximum charge of the patterned beam was measured to be approximately $\sim 15$~nC corresponding 
to an average charge of   $\sim (15\mbox{~ nC})/(8\times8) \simeq 200$~pC per beamlet.

\begin{figure}
\centering
\includegraphics[width=1.0\linewidth]{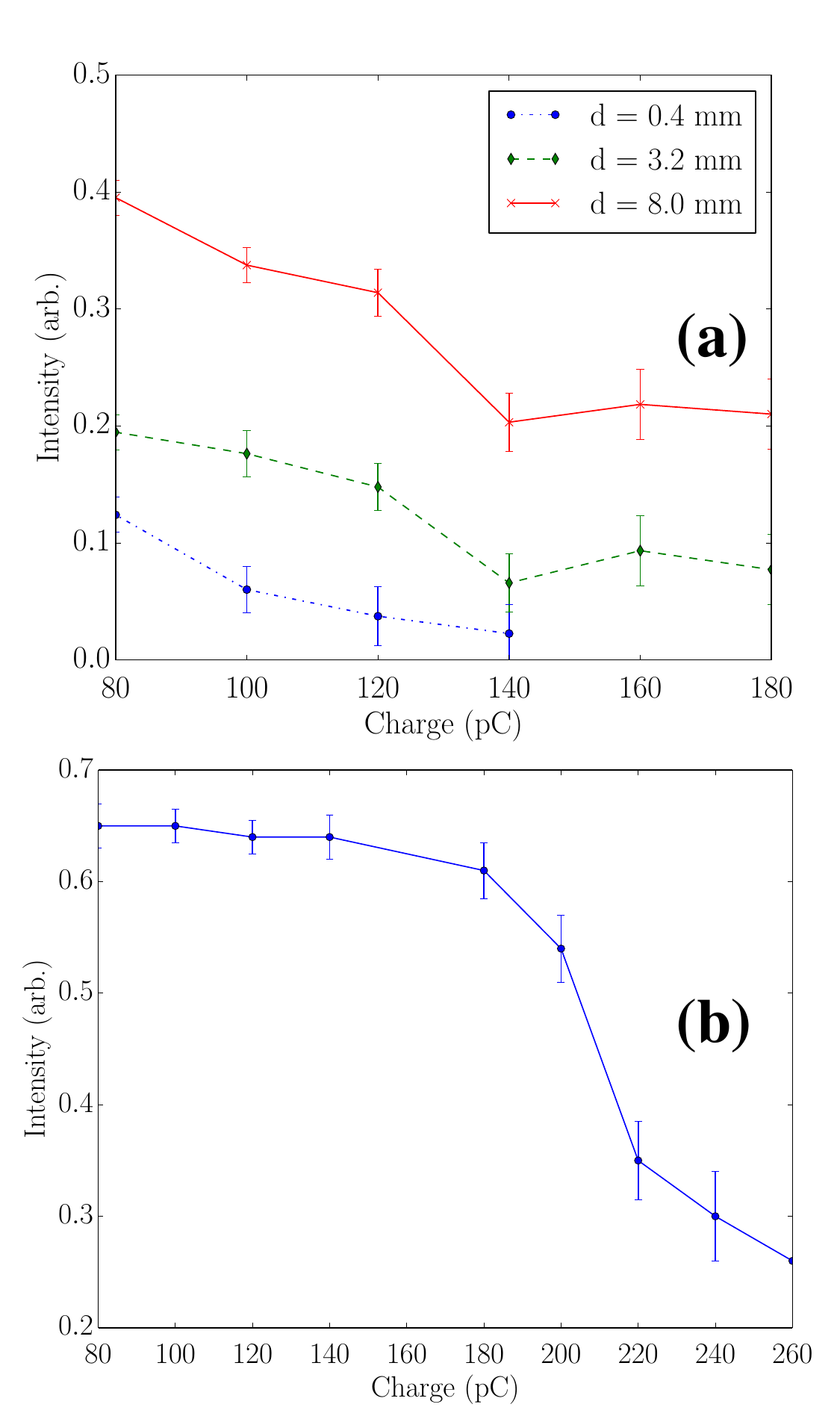}
\caption{\label{ebeammod}
Transverse bunching factor evaluated at its lowest-frequency maximum versus bunch 
charge for the three cases of solenoid settings displayed in Fig.~\ref{YAG1} with 
corresponding beamlet spacing $d$ (a) and for the case of a solenoid field of 290~A with associated beamlet spacing of $d=10$~mm (b).}
\end{figure}

\begin{figure}
\centering
\includegraphics[width=1.0\linewidth]{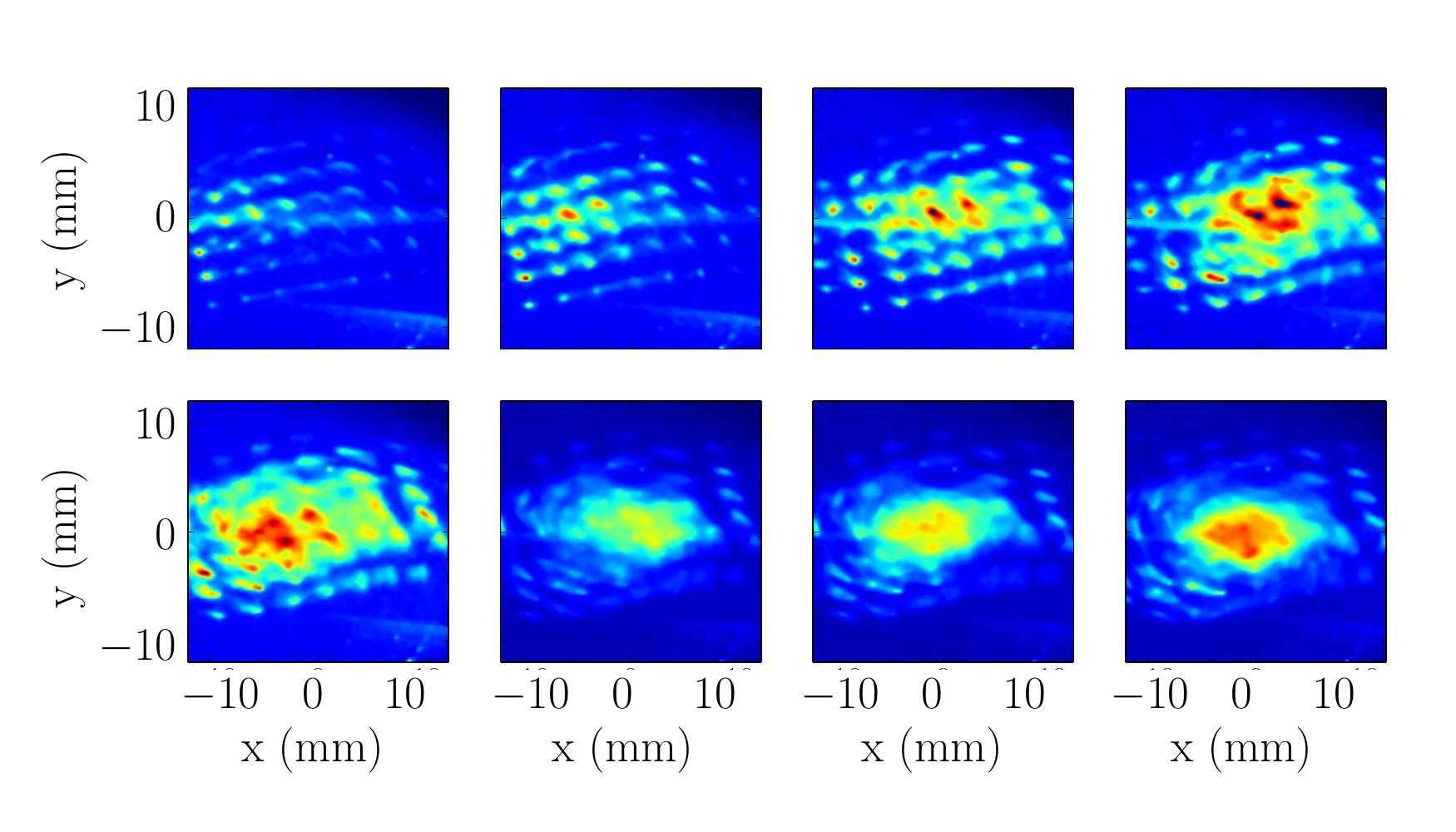}
\caption{\label{YAG1b}False color measured 48 MeV electron beam patterns for various charges. From left to right and top to bottom: $Q=60$, 100,
 200, 300, 400, 500, 600, 700~pC with a matching-solenoid current of 240~A. }
\end{figure}
The resulting electron beamlet formations pictured on Fig.~\ref{YAG1} were 
analyzed using the same Fourier analysis as the one used in Section~\ref{sec:opticalanalysis} for the UV-laser images. 
Figure~\ref{ebeammod} (a) summarizes the evolution of the transverse bunching factor at its lowest-frequency maximum
versus total bunch charge for different matching solenoid settings. The average beamlet separation $d$
changes from 0.4 mm to 8 mm at YAG1 location with the matching solenoid current increased from 215 A to 270 A.
Figure~\ref{ebeammod} (b) gives the evolution
of the transverse bunching factor for the case of $d=10 \pm 0.4$ mm. One can see the modulation
is fully determined by solenoid imaging at charges of $Q<180$ pC.

Finally, it should be noted, that the measurements were taken at YAG1 and do not provide information on 
possible modulation reappearance at a downstream position along the beamline: the betatron 
phase advance at a downstream observation point could be such that the modulation
is washed out in the position space but prominent in the 
angular coordinate.

\subsection{Multi-beam acceleration to 48~MeV}
The modulation introduced on the cathode propagated and preserved 
through the beamline up to the transverse-to-longitudinal 
emittance-exchange (EEX) beamline entrance; see Fig.~\ref{YAG1b}.
 There should be no strong focusing applied along the low-energy beamline 
 as close encounter of the beamlets produces
 strong distortion as explored in Ref.~\cite{Rihaoui}. 
Consequently, the low-energy beamline  elements should be 
properly matched to allow the large waist. 
At medium energy, the transverse space-charge force is
significantly decreased and therefore not expected to impact the multi-beam dynamics.
In order to avoid a tight waist at low energy we used 
the linac solenoid LS1 (see Fig.~\ref{beamline}) to image 
the beamlet pattern directly on the YAG5 screen located 14~m downstream
of the photocathode surface and just prior to the EEX beamline. 
At this location, the beam energy is measured to be 48 MeV.
Figure~\ref{YAG1b} shows the beam distribution at YAG5 for different bunch charge. 
The typical beamlets separation (center to center)
is on the order of $\sim 3$~mm $\pm 0.3$~mm. Such a distribution could be 
further manipulated using a telescope composed of four quadrupole magnets to generate 
a train of short bunches along the temporal axis downstream of the EEX beamline~\cite{PhysRevLett.118.104801}.
Such bunch train could possibly support the generation
of THz radiation using, e.g., coherent transition radiation, or the resonant excitation
of wakefields in a wakefield
structured such as a dielectric-lined waveguide~\cite{Qiang:NAPAC16}. 
As already noted, the coupling at YAG5 could be removed by mounting the MLA assembly on a 
rotatable mount. Likewise, the coupling could be taken advantage of to select an 
angle so that a smaller projected separation along the horizontal axis could be achieved. 
Such a configuration would provide a knob to continuously
vary the beamlets separation (e.g. and THz-enhancement frequency) downstream of the EEX beamline. 

\section{Generation of magnetized multi-beams}

In this section, we describe a possible applications of the patterned electron 
beam form by the MLA setup as a beam-based diagnostic tool for inferring the
residual axial magnetic field at the photocathode plate by measuring the value of canonical angular momentum (CAM).

\begin{figure} [t]
 \includegraphics[width=0.99\linewidth]{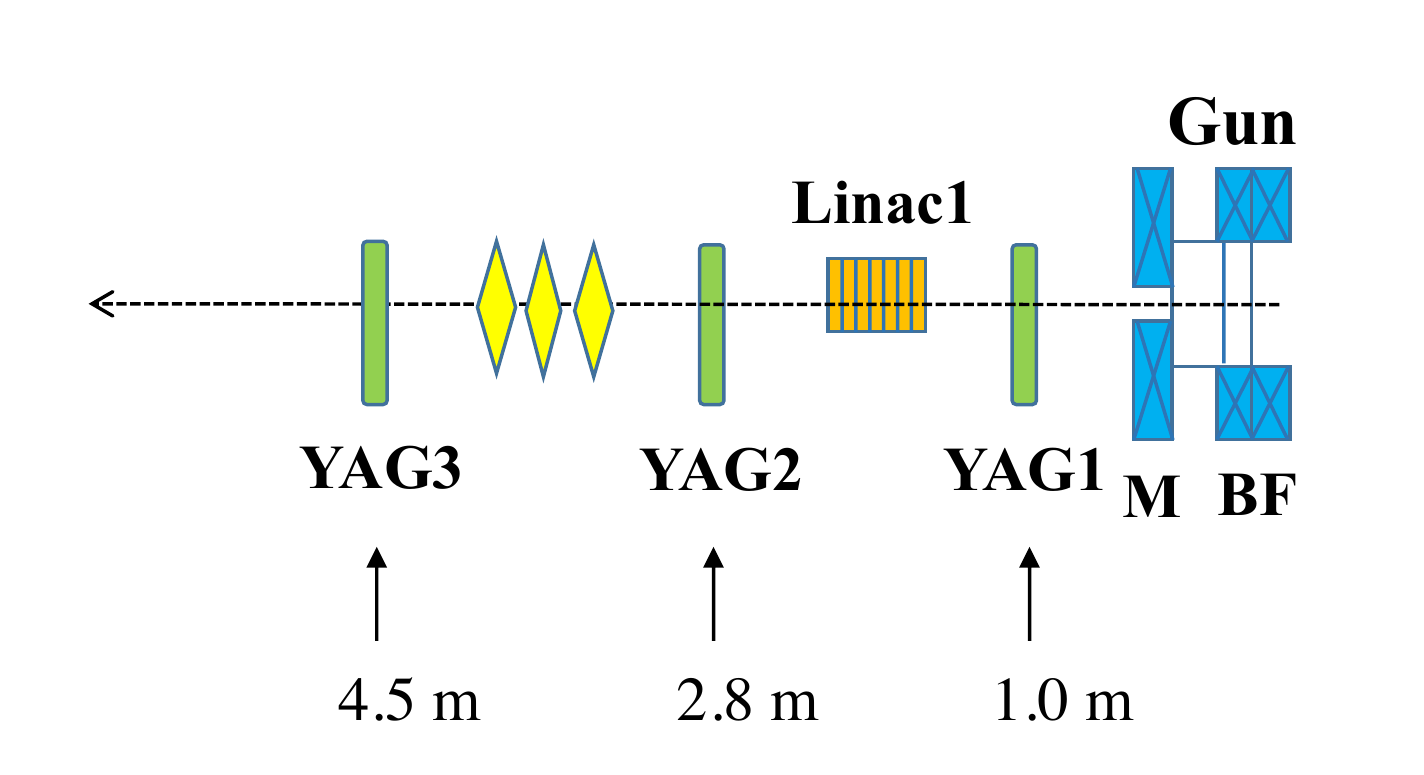}
\caption{\label{awarfgun} Overview of the AWA-WB beamline. The bucking (B) and focusing (F) solenoidal lenses can be 
setup to produce CAM-dominated beams. The positions of the YAG viewers are denoted in meters.
}
\end{figure}

As an example we consider the Argonne Wakefield Accelerator ``witness-beam"  (AWA-WB) beamline diagrammed in Fig.~\ref{awarfgun}. 
In brief, AWA-WB accelerator incorporates an L-band $1+\frac{1}{2}$ RF gun with a Mg photocathode on its back plate. 
The gun is surrounded by a bucking and focusing solenoids, nominally powered to yield a vanishing magnetic field $B_{0z}$ at the photocathode surface. 
The solenoids can be tuned and provide a non-vanishing axial magnetic field $B_{0z}$ at the cathode. 
Nominally, the bucking and focusing solenoids have the opposite polarity, however
they can be operated with the same polarity and provide significant field ($B_{0z}\sim 0.1$~T)  on the photocathode. 

\subsection{Magnetized beams}
According to Busch's theorem the total canonical angular momentum of an electron in an cylindrical-symmetric magnetic field  is 
conserved and given by \cite{Reiser}
\begin{equation}\label{busch}
 L = \gamma m r^2 \dot \theta+\frac{1}{2}e B_z(z)r^2 + \mathcal{O}(r^{4}),
\end{equation}
where ($r,\theta,z$) refers to the electron transverse position the cylindrical coordinate system. 
When a electron beam is born in presence of an axial magnetic field, it forms a ``magnetized'' beam state. 
Such beams have a variety of applications in electron cooling and can be further manipulated to form  
beams with asymmetric transverse emittances or ``flat'' beams \cite{Brinkmann:2000jy}.

 The conservation of the CAM $L$ from Eq.~\ref{busch} yields 
the mechanical angular momentum (MAM) of the beam in the magnetic-field-free zone to be
\begin{equation}
\label{magnetization}
 |\pmb L| = \gamma m |\pmb r \times \frac{d\pmb r}{dt}| = \frac{1}{2}e B_{0z} r_{0}^2,
\end{equation}
where $B_{0z} $ is the field at the cathode surface, $r_0$ and $r$ are 
respectively the electron radial  coordinate on the photocathode surface
and at a downstream magnetic-field-free location. The norm of $L\equiv |\pmb L|$ can be computed as $L=|\pmb r \times \pmb p|=|x p_y - y p_x|$.
Following Ref.~\cite{Kim:2003cp}, we characterize 
a CAM-dominated beam via its magnetization $\mathcal{L}\equiv\langle L\rangle/2\gamma m c$ where $\mean{L}= e B_{0z} \sigma_{0}^2$
represents the statistical averaging of $L$ over the beam transverse distribution and $\sigma_0$ the RMS 
transverse radius of the electron beam on the photocathode surface.

\subsection{Method to measure $\cal L$}

\begin{figure}
 \includegraphics[width=0.99\linewidth]{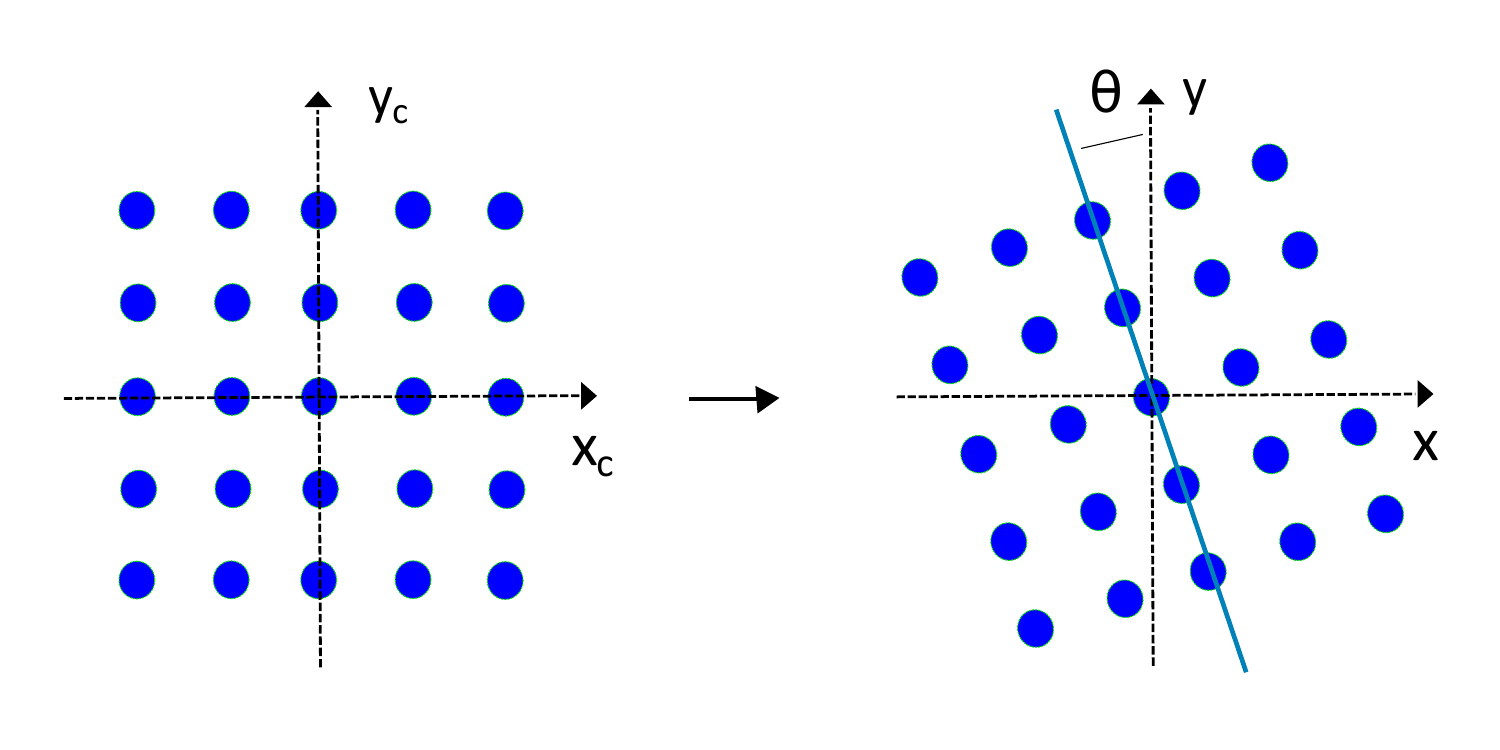}
\caption{\label{sketch}Illustration of the similarity transformation between the 
initial beamlet pattern emitted from the phtocathode (left)as it propagates to a downstream location (right).
This schematics assumes the transverse momentum is solely angular in an axisymmetric external-focusing lattice.}
\end{figure}

We now consider the multi-beam laser distribution discussed in Section \ref{multibeam} impinging a photocathode immersed in an 
axial magnetic field. The resulting electron beam, composed of multiple beamlets,
is born with a CAM and will therefore undergo a similarity transformation (in the presence of a axisymmetric external focusing) of the form 
\begin{eqnarray}~\label{eq:Pexp2}
\begin{pmatrix} x  \\  y \end{pmatrix} &=& [ k +R(\theta) ]  \begin{pmatrix} x_c \\  y_c  \end{pmatrix},
\end{eqnarray}
after exiting the magnetic-field region; see Fig.~\ref{sketch}. In the previous equation 
the subscript $_c$ corresponds to the spatial coordinates on the cathode surface, $k$ is a scalar and $R(\theta)$ is the $2\times2$ matrix 
associated to a rotation with angle $\theta$. 
 
A measurement of the rotation angle and array size provides the value of mechanical angular momentum.
\footnote{A similar method was discussed in Ref.~\cite{Sun:2004zu} for 
the case of slits located downstream of the electron source.} Such a measurement relies on the measurement of the pattern evolution between
two axial locations. Considering two YAG:Ce screens separated by a drift space with length $D$, 
the mechanical angular momentum $L$ can then be deduced by computing the relative 
rotation of the beamlet pattern $\theta$ between the two locations as~\cite{Sun:2004zu}
\begin{equation}
\label{mag2}
 L = \frac{p_z}{D}\left[\left(\frac{n}{2}a_1\right)\right]^2(M \sin{\theta}),
\end{equation}
where $p_z$ is the axial momentum, $n$ is a number of beamlets, $a_1$ is the separation between beamlets at the first viewer,
and $M=a_2/a_1$ is the magnification factor between second and first viewer. Relating it to Eq.~\ref{magnetization}
one can infer the value of the magnetic field on the cathode $B_{0z}$.
\begin{figure}
 \includegraphics[width=1\linewidth]{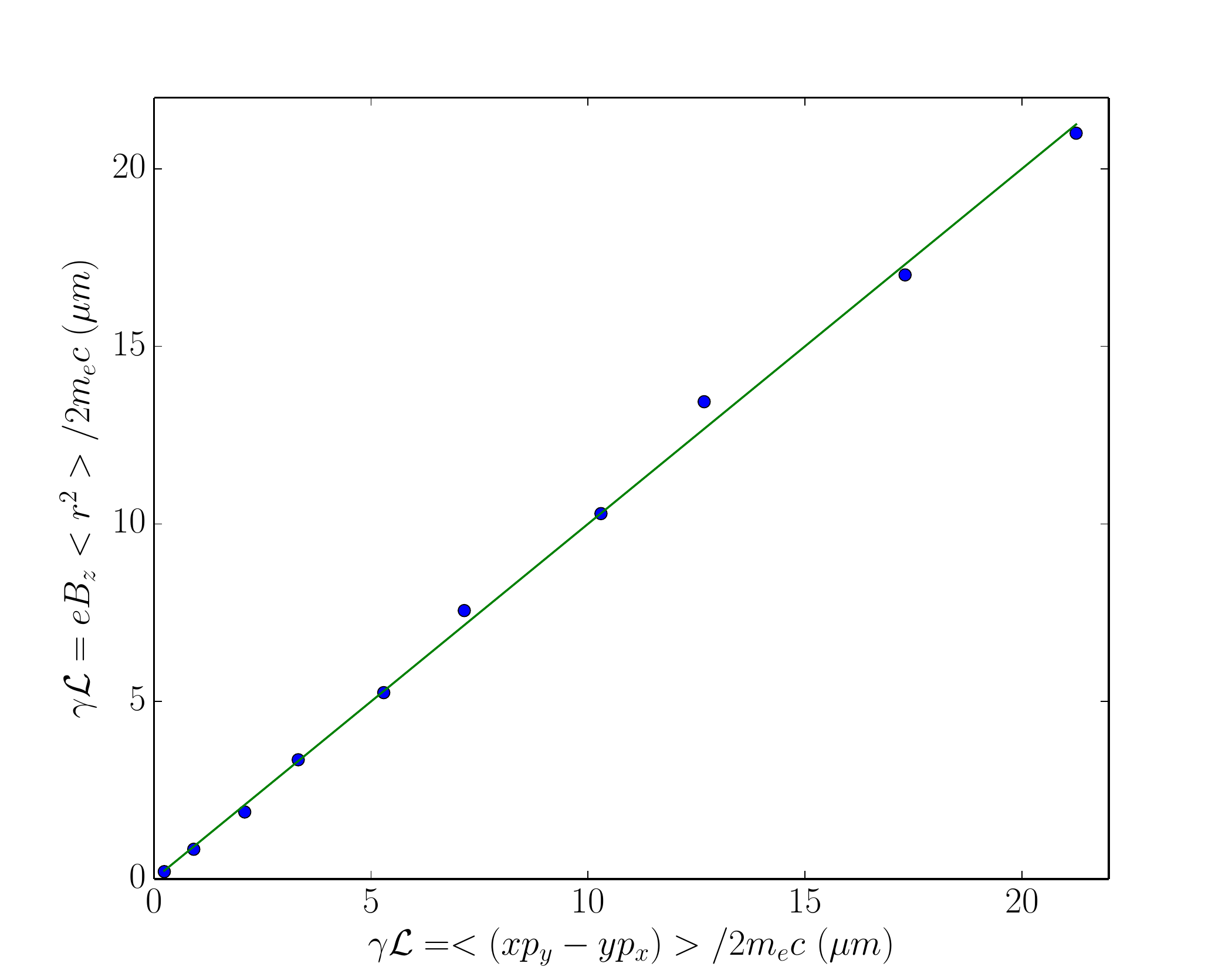}
\caption{\label{lplot}Proof-of-principle via numerical simulations using {\sc Impact-T} of the measurement method based on Eq.~\ref{mag2}. 
The magnetization $\mathcal{L}$ is computed from the particle distribution
(blue markers) and plotted against the magnetization inferred from the $B_{0z}$ on the cathode surface. The green solid line correspond to the diagonal line.}
\end{figure}

To validate the proposed method via numerical 
simulations we used {\sc impact-t}~\cite{ImpactT} and considered the AWA-WB beamline diagrammed in Fig.~\ref{awarfgun}.
Laser multi-beam array was converted into {\sc Impact-T} particle distribution
using the method described in Section \ref{homogenizer}. It was then propagated
through the beamline and 
saved at two locations 
of YAG screens (YAG2 and YAG3 in Fig.~\ref{awarfgun}) and away from the waist.
The centroids of each beamlet were found and the mechanical angular momentum is inferred from  
Eq.~\ref{mag2}. We performed numerical simulations for different $B_{0z}$ field values and results 
are summarized in Fig.~\ref{lplot}.

The latter Figure confirms that the CAM (as inferred from the value of $B_{0z}$) 
is fully transferred to the MAM. Some systematic discrepancies ($<5\%$) are observed as $\mathcal{L}$ 
increases and most likely due to the contribution of nonlinear terms in multipole expansion of $B(z)$
not accounted for in Eq.~\ref{busch} (which assumes a paraxial linear approximation). 
Likewise, space-charge effects might alter the results.
Both limitations will be the object of further studies. 
Nevertheless, the simulations demonstrate that illuminated
cathode with a patterned laser beam provides a simple method to measure the MAM. 
It should be noted, that the fundamental systematic error may also come from
diagnostic cameras mutual misalignment. In a case of low $\mathcal{L}$, a longer
distance between two screens has to be employed.
Additionally, this technique provides an excellent determination of the magnetic axis and probes the laser spot alignment. 

To extract the rotation angle and the beamlet separation, one can calculate
beamlet positions using conventional 2D peak finding algorithm. 
However, this method becomes not very robust when the beamlet formation size at the 
location of the second screen is bigger than the screen size.

\begin{figure}[hhhh!!!!]
 \includegraphics[width=1\linewidth]{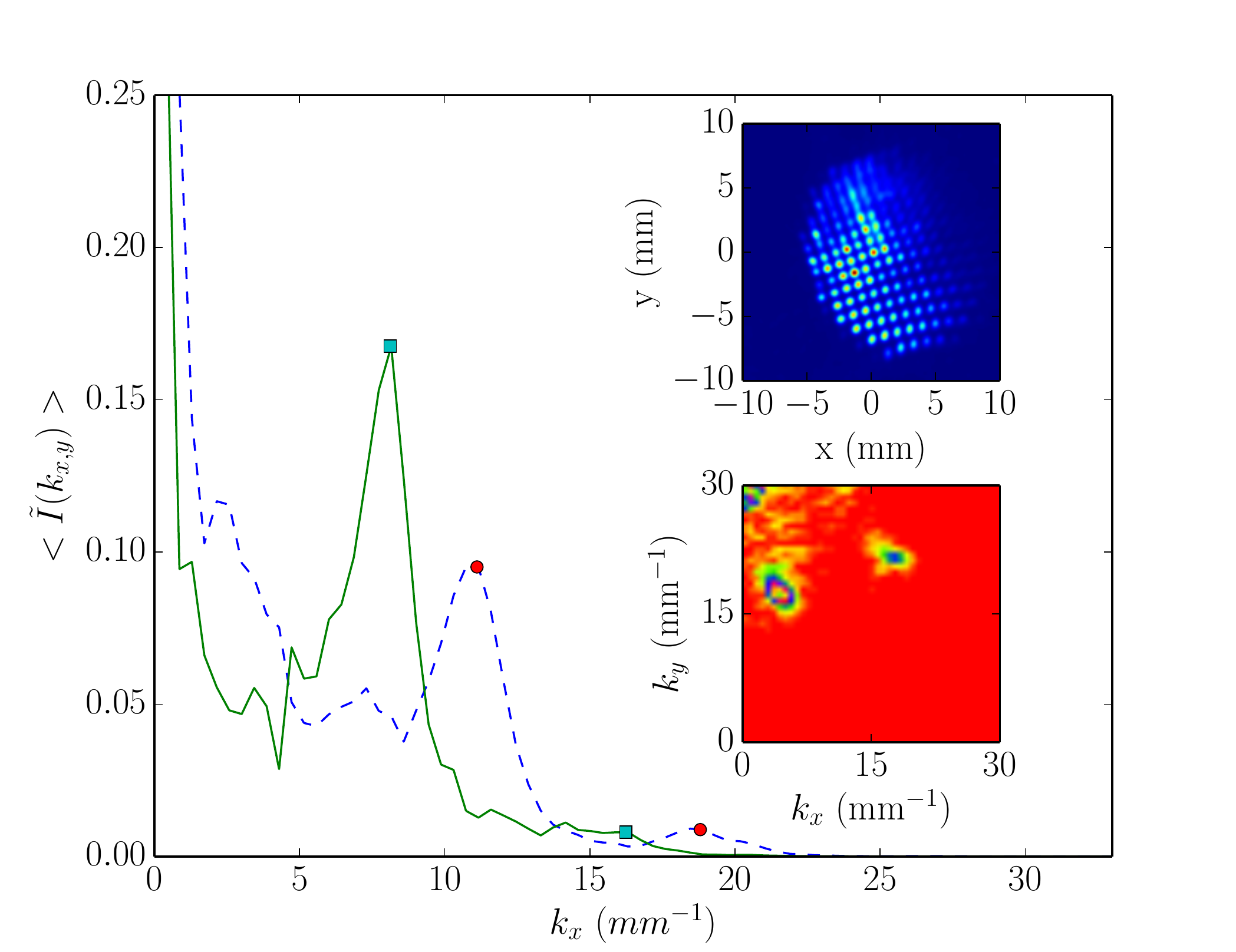}
\caption{\label{fftanalysis} Bunching factor as function 
of horizontal $k_x$ (solid trace) and vertical $k_y$  (dashed trace) spatial frequency. 
The ratio of the lowest-frequency peaks $k_{y1}/k_{x1}$ provides a measurement 
of the pattern rotation angle with respect to the horizontal axis. This data is extracted from the 
corresponding beam image (upper inset) and from projections of its 2D FFT image (lower inset).}
\end{figure}

Another approach is to utilize projections $\mean{\tilde{I}(k_{x})}$ and $\mean{\tilde{I}(k_{y})}$
of the images 
in reciprocal Fourier space calculated via 2D fast Fourier transform (FFT). 
In this case, the tilt of the image will result in a difference between the locations of first harmonics; see Fig.~\ref{fftanalysis}.
The tilt angle can be then computed as $\tan \theta = k_{x_1}/k_{y_1}$,
where $k_{{(x,y)}_{1}}$ is the coordinate of the first-harmonic peak of the spatial bunching factor along the corresponding axis. 
Such an analysis assumes the beamlet-pattern periodicity along the two directions is identical (which is the case in our experiment).
The full-width half-maximum (FWHM) sizes of the peaks in Fig.~\ref{fftanalysis} are then accounted as errorbars of the measurement. 

\subsection{Electron beam experiment}
Proof-of-principle electron beam experiment was performed at AWA-WBA beamline.
 A $12\times 12$ laser beamlet pattern with
 rms duration of 6~ps was formed by using the technique from Section \ref{multibeam}.
The $\sim 5$-MeV beam out of the RF gun was further accelerated using the L-band linac to $\sim 10$~MeV; 
see Fig. \ref{awarfgun}. In the experiment, the total charge was 60~pC per bunch, resulting in $\sim 420$~fC per beamlet. 

 \begin{figure}[hhhhh!!!!!!]
 \includegraphics[width=1\linewidth]{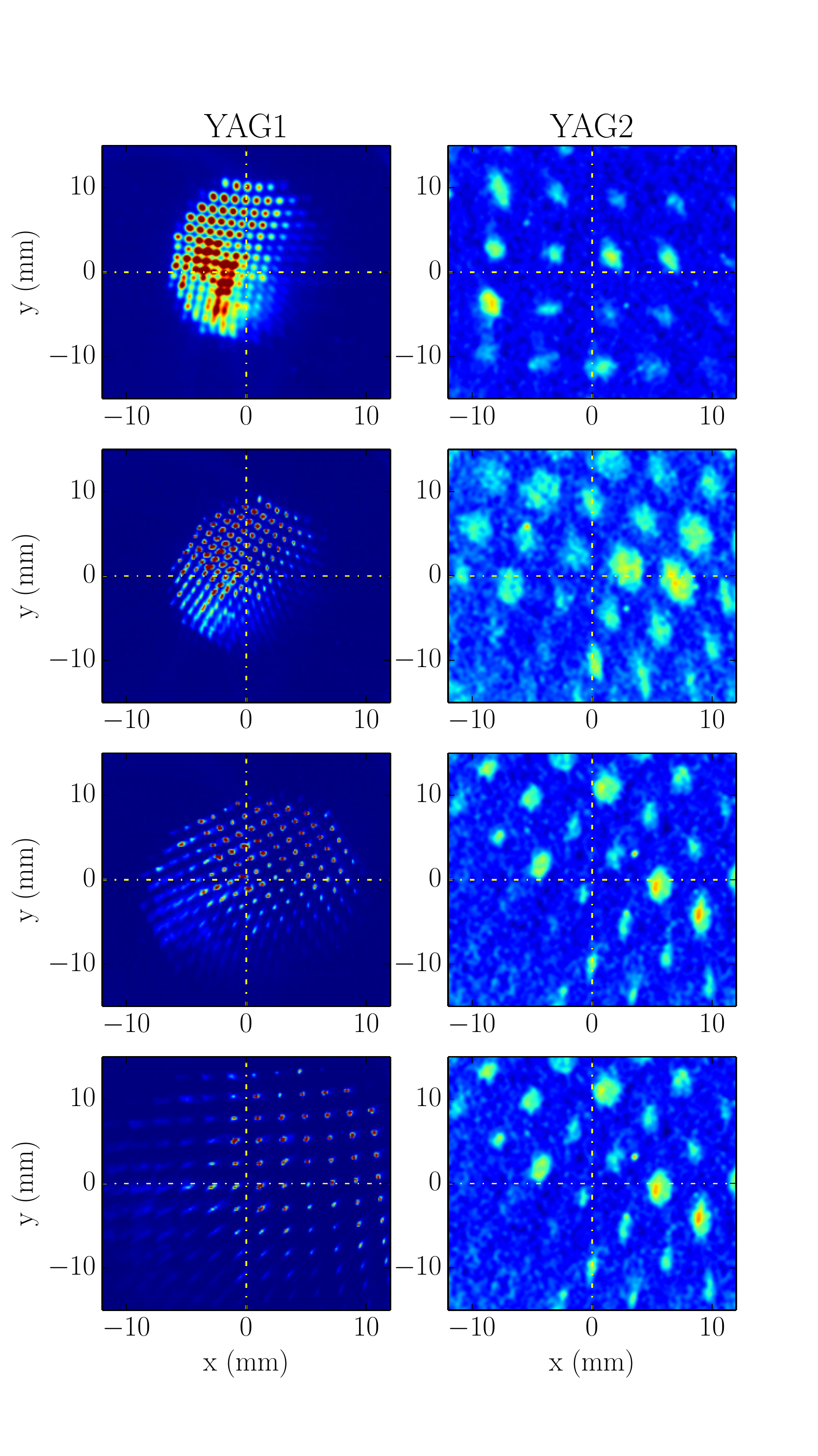}
\caption{\label{twoyag} Beamlet-pattern distribution on YAG1 (left column) and YAG2 (right column)
for different settings of the magnetic field on the photocathode $B_{0z}=0$, 200, 500, 1000~G (from top to bottom).  
The  radial momentum (compared to the assumption made in the diagram of Fig.~\ref{sketch}) yields to 
substantial magnification at YAG2 in addition to the rotation.}
\end{figure}

The three solenoids depicted in Fig.~\ref{awarfgun} were controlled independently via unipolar power supplies.
 We started with the normal operational configuration where the bucking and focusing 
solenoids had opposite polarities which yields relatively
 low magnetization of the beam.
 The bucking solenoid current was slowly decreased to 0 A and the
 induced rotation of the beamlet formation was observed at YAG1 and 
 YAG2 locations; see Fig.~\ref{twoyag}.
Then the polarity of the bucking and focusing solenoids was flipped and the bucking solenoid current was ramped up to -500A.
 Total of 20 bucking solenoid current values were used to reach the maximum field at the cathode surface
of $\simeq1400$~G.

\begin{figure}[hhhh!!!!!!!!!]
 \includegraphics[width=.95\linewidth]{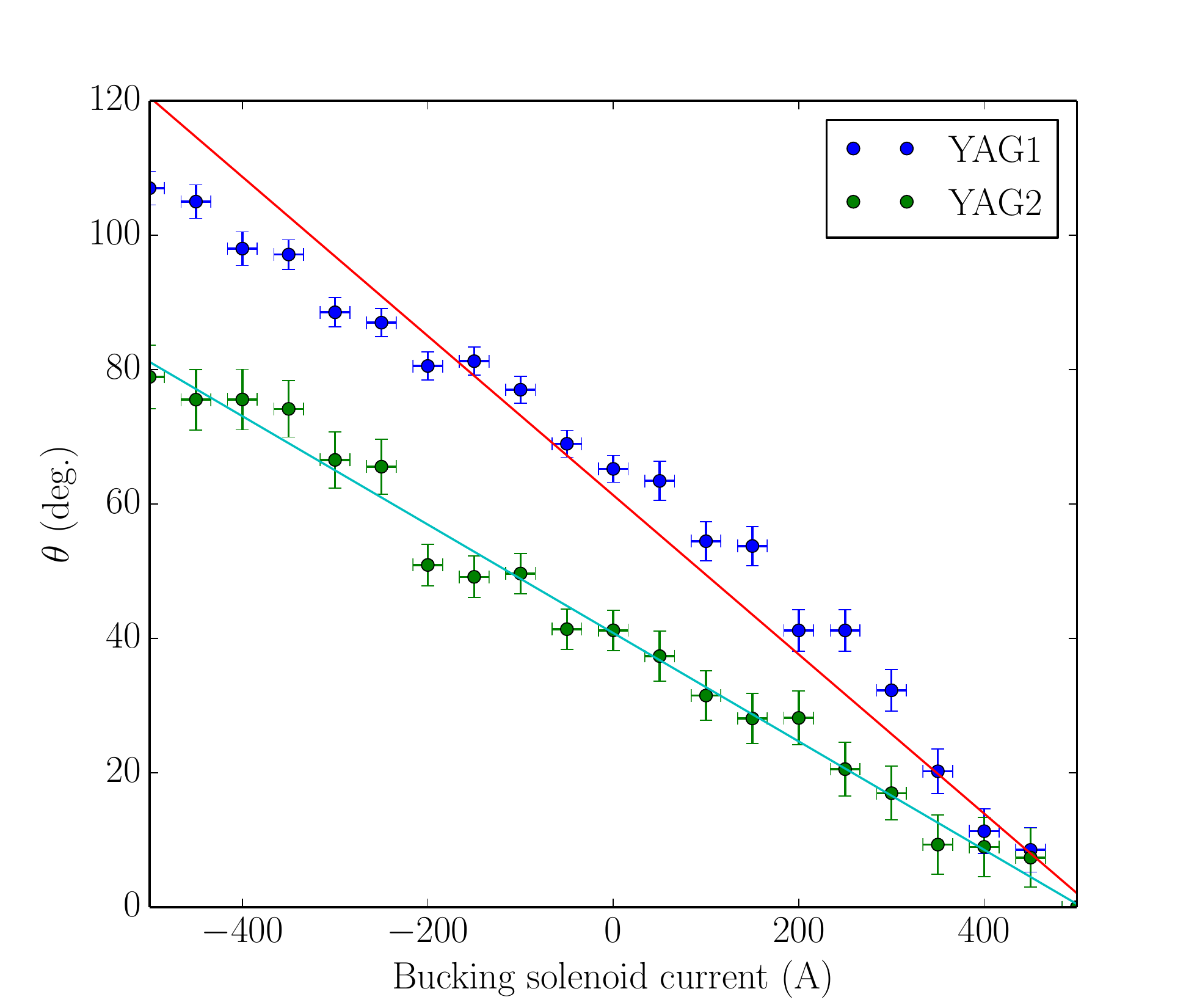}
\caption{\label{angles}Rotation angle of the beamlet pattern with respect to the vertical angle at screens YAG1 (left) and YAG2 (right) as function of 
the bucking-solenoid magnetic field. The lines correspond to a linear regression of the experimental data.}
\end{figure}

The beamlet pattern at the two screens YAG1 and YAG2 makes a different rotation angle. The rotation of the pattern between the two screens $\theta$ increases
with the magnetization, as it can be seen in Fig.~\ref{angles}. In the latter picture the data point
were obtained from the 2D FFT technique detailed in the previous Section. From the inferred rotation angle $\theta$, the MAM was recovered via Eq.~\ref{mag2}.  

\begin{figure}[hhhh!!!!!!]
 \includegraphics[width=.95\linewidth]{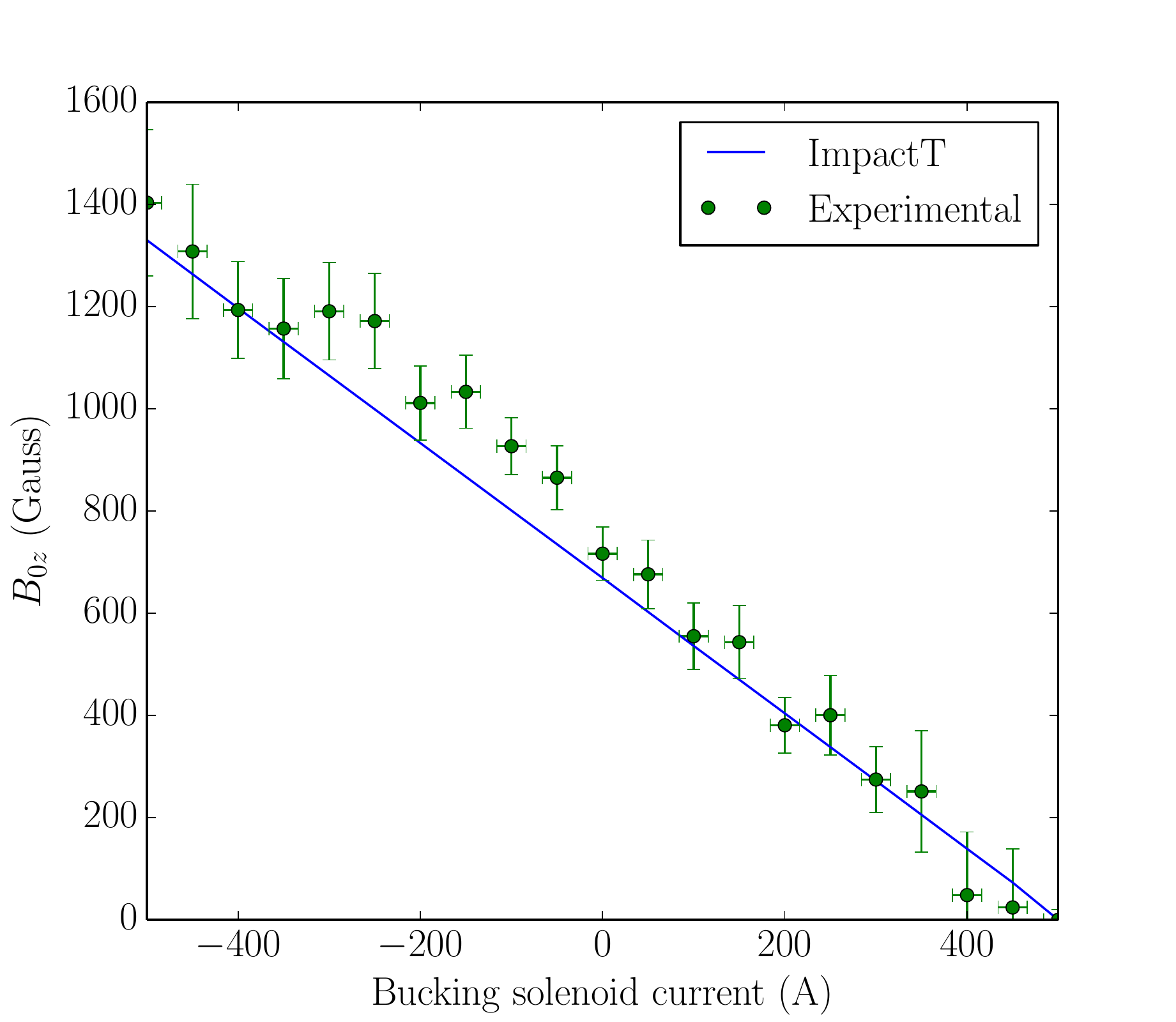}
\caption{\label{experiment2}Comparison between {\sc Impact-T} simulations (solid blue line) and experimentally recovered values
of $B_{0z}$ (symbols with error bars) from Fig.~\ref{twoyag}.}
\end{figure}

The retrieved value of the $B_{0z}$, the applied field on the cathode surface, computed using the data of Fig.~\ref{angles} 
for different currents of the bucking solenoids are reported in Fig.~\ref{experiment2}. Specifically, the retrieved values are computed via 
$B_{z0}=\frac{2 m_e c^2 L}{ec}r^2$ where $L$ is found from Eq.~\ref{mag2}. These 
values are in very good agreement with {sc impact-t} simulations of the measurement which includes a model of the
solenoids simulated with {\sc poisson}~\cite{poisson}.

\section{Summary}
We demonstrated the possible use of a microlens array to control the transverse distribution of a photocathode laser pulse and associated 
photoemitted electron bunches. We especially confirm that this simple and rather inexpensive setup 
could be used to homogenize the beam transverse distribution thereby improving the transverse emittance. 
Additionally, we investigated the generation of patterned electron beams consisting of multiple transversely-separated beamlets. 
The latter type of beam could be used for various application and could yield temporally-modulated electron beam when combined
with a transverse-to-longitudinal emittance-exchange beamline but also serve as a beam-based diagnostics. 
We illustrated the application of the patterned beam to diagnose the magnetization of a magnetized beam 
(by using the beam evolution to infer the the axial magnetic applied at the photocathode surface). The application of patterned beams could be
further extended to explore, e.g, nonlinearities in linear
accelerators and characterize beamline element (transfer matrix measurement). Given its simplicity, 
low cost and versatility we expect the present work to motivate further applications to photoemission electron sources and laser-heater system.

%In such a study each beamlet can be viewed as a macroparticle and the relative displacement 
%of the macropaticles (or distortion of the array) could provide some 
%information on the relative transverse force experienced by the macroparticles. 

\section{Acknowledgements}
This work was supported by the US Department of Energy under contract No. DE-SC0011831 with Northern Illinois University. 
The work by the AWA group is funded through the U.S. Department of Energy, Office of Science, under contract No. DE-AC02-06CH11357.
The work of A.H. and partially P.P. is supported by the US Department of Energy under contract No. DE-AC02-07CH11359 with Fermi Research Alliance, LLC.

\end{document}